\documentclass[10pt, twocolumn, journal, final]{IEEEtran}          
\IEEEoverridecommandlockouts
\usepackage{amsmath}

\usepackage{examplep}  
\usepackage{graphicx}
\usepackage{cite}
\usepackage{algorithmicx}
\usepackage[ruled]{algorithm}
\usepackage{algpseudocode}
\usepackage{mathtools}
\usepackage{dsfont}
\usepackage{amsfonts}
\usepackage{amssymb}
\usepackage{balance}
\usepackage{booktabs,tabularx}
\usepackage{multirow}
\usepackage{textcomp}
\usepackage[usestackEOL]{stackengine}
\usepackage[hyphens]{url}
\usepackage{pict2e}
\usepackage{color}
\usepackage{enumerate}
\usepackage{float}
\usepackage{mathrsfs}
\usepackage{mathtools}
\usepackage{tikz}
\usetikzlibrary{shapes.arrows,decorations.pathreplacing,decorations.markings,shapes.geometric}
\tikzset{radiation/.style={{decorate,decoration={expanding waves,angle=0,segment length=2pt}}}} 
\usetikzlibrary{positioning}
\usepackage{graphicx}
\usepackage[normalem]{ulem}
\newcommand{\trusted}[1]{\text{sup}(#1)}
\newcommand{\unavailable}{\triangleleft ||}
\newcommand{\believes}{\mid\!\equiv}
\newcommand{\oncesaid}[1]{{\overset{{{ {#1}}}}{\mid\!\sim}}}

\newcommand{\fresh}[1]{\#(#1)}

\newcommand{\sees}[1]{{\overset{{{ { #1 }}}}{ \triangleleft }}}

\newcommand{\sharekey}[1]{{\overset{{{ {#1}}}}{\leftrightarrow}}}
\newcommand{\isitequal}{\overset{\scalebox{0.6}{\textbf{?}}}{=}}

\DeclareMathAlphabet\mathbfcal{OMS}{cmsy}{b}{n}

\makeatletter
\def\blfootnote{\gdef\@thefnmark{}\@footnotetext}
\makeatother

\begin{document}
\title{Multi-factor Physical Layer Security Authentication in Short Blocklength Communications}
\author{Miroslav Mitev, \and Mahdi Shakiba-Herfeh,~\IEEEmembership{Member,~IEEE},         \and
        Arsenia Chorti,~\IEEEmembership{Senior Member,~IEEE},        \and
        Martin Reed,~\IEEEmembership{Member,~IEEE}        
      \thanks{ M. Mitev, M. Shakiba-Herfeh and A. Chorti are with ETIS UMR8051, CY Cergy Paris University, ENSEA, CNRS, F-95000, Cergy, France (\{miroslav.mitev, mahdi.shakiba-herfeh, arsenia.chorti\}@ensea.fr); }
      \thanks{M. Reed is with the CSEE, University of Essex, Colchester (mjreed@essex.ac.uk), UK;}
      \thanks{M. Mitev is supported by the DIM RFSI project SAFEST; M. Shakiba-Herfeh and A. Chorti are supported by the ELIOT ANR-18-CE40-0030 - FAPESP 2018/12579-7 and the INEX Project eNiGMA; M. Reed  is supported by the H2020 project SerIoT with project agreement no 780139 funded by the EU.}
}
\maketitle
\begin{abstract} 


Lightweight and low latency security schemes at the physical layer that have recently attracted a lot of attention include: (i) physical unclonable functions (PUFs), (ii) localization based authentication, and, (iii) secret key generation (SKG) from wireless fading coefficients. In this paper, we focus on short blocklengths and propose a fast, privacy preserving, multi-factor authentication protocol that uniquely combines PUFs, proximity estimation and SKG. We focus on delay constrained applications and demonstrate the performance of the SKG scheme in the short blocklength by providing a numerical comparison of three families of channel codes, including half rate low density parity check codes (LDPC), Bose Chaudhuri Hocquenghem (BCH), and, Polar Slepian Wolf codes for $n=512, 1024$. The SKG keys are incorporated in a zero-round-trip-time resumption protocol for fast re-authentication. All schemes of the proposed mutual authentication protocol are shown to be secure through formal proofs using Burrows, Abadi and Needham (BAN) and Mao and Boyd (MB) logic as well as the Tamarin-prover.

\end{abstract}

\section{Introduction}
\label{sec:intro}

Authentication is central in building secure Internet of things (IoT) networks; confirming the identity of devices and their role in the network hierarchy eliminates the possibility of numerous attacks \cite{iot_security_survey,shakibaPLS, Jun_iot2020}. However, the low latency and computational power constraints present in many IoT systems \cite{Chorti_sensors}, render the design of IoT authentication mechanisms a challenging task. In this direction, a 3GPP report on the security of ultra reliable low latency communication (URLLC) systems notes that authentication for URLLC is still an open problem~\cite{3gppURLLC}. Current solutions rely on modulo arithmetic in large fields and typically incur considerable latency, in the order of tens of milliseconds, exceeding the delays that are tolerated in applications such as vehicle-to-everything (V2X)~\cite{Teniou18}.
Moreover, with the advance of quantum computing, traditional asymmetric key cryptographic schemes will become semantically insecure while, at the same time, current proposals for post-quantum alternatives use keys of impractical lengths \cite{asiacrypt-2017-29693}. Therefore, the proposal of new, lightweight, security primitives and protocols for device authentication is timely, especially in the low latency and IoT context. 

In this sense, physical layer security (PLS) has proven itself as a lightweight alternative to computational complexity based schemes \cite{A_Chorti_PLS0, A_Chorti_PLS1}. The increasing interest in PLS has been stimulated by many practical needs. Notably, many critical IoT networks require fast authentication, e.g., in V2X applications, telemedicine and haptics. Moreover, PLS, that relies upon information-theoretic security proofs, could resist quantum computers, unlike corresponding asymmetric key schemes relying on the intractability in polynomial time of certain algebraic problems. 

PLS schemes exploit physical layer entropy sources, including  both in the hardware, as well as in the communication medium~\cite{Eurasip2020,Mitev_sub-scheduling, Globecom_2019_MiM}. With respect to the former, physical unclonable functions (PUFs) are hardware entities harnessing entropy from physically unclonable variations that occur during the production process of silicon~\cite{First_silicon_puf,PUF}. Due to their unclonability, PUFs can be used in challenge -- response authentication protocols, where a challenge can refer to measuring the jitter of a ring oscillator, power-on state, etc., ~\cite{PUF_Location, Auth_protocol_pre-shared-key, Data_provenance_puf, Auth_protocol-sess_key-from_nonces1}. 


\begin{table*}
\caption{Contributions of this paper.}
\label{table:contributions}
\begin{tikzpicture}[overlay]
\draw[rounded corners=10pt,line width=0.5pt, densely dashed] (0.5,-1)--(11.3,-1)--(11.3,2.25)--(0,2.25)--(0,-1)--(.5,-1);
\draw[rounded corners=10pt,line width=0.5pt, densely dashed] (0.5,-2.54)--(11.3,-2.54)--(11.3,-1.2)--(0,-1.2)--(0,-2.54)--(.5,-2.54);
\node [rotate=90] at (0.4,0.8) {Initial authentication};
\node [rotate=90] at (0.5,-1.85) {\Longstack[c]{0-RTT \\ resumption}};
\end{tikzpicture}
\begin{minipage}[t]{0.05\textwidth} 
${}$ 
\end{minipage}
\begin{tabular}{ |p{2.4cm}|p{7cm}|p{6cm}|}
 \hline
Technology & Novelty & Overall contributions \\ \hline
${}$\newline Proximity estimation
\newline \newline \newline  \newline  PUFs 
\newline \newline\newline\newline SKG \newline\newline \newline\newline   SKG  &
As an initial factor of authentication we propose a novel proximity detection mechanism leveraging user mobility to authenticate a static server.
\newline
Standard PUF mechanisms are used as a main factor of authentication in our two-party protocol. Combining PUFs and mobility-based proximity estimation can prevent impersonation attacks in the presence of a malicious server.
\newline
The proposed system exploits the reciprocity of the wireless fading coefficients between two terminals to generate maximum entropy session keys using short blocklength encoders. 
\newline  ${}$ \newline
In a subsequent communication between the nodes the generated keys are incorporated in a novel PHY-based 0-RTT protocol to provide forward secrecy and protection against replay attacks.
 & 
 \vspace{0.05cm}
The combination of all the technologies in a single solution gives a novel and secure authentication protocol.
\newline
To support the employment of PHY layer SKG in our short blocklength communication protocol we provide a numerical comparison of three families of Slepian Wolf reconciliation codes.
\newline
We validate the performance of the  proposed mobility-based proximity detection through real-life experiments.
\newline
The security properties of the protocol are formally proven using MB logic and the Tamarin-prover.
\newline
We introduce a novel, physical layer, forward secure 0-RTT resumption authentication mechanism.\\
\hline          
\end{tabular}
\end{table*}

With respect to PLS techniques using the communication medium, these include secret key generation (SKG) from shared randomness and localization. In fact, it is well established that SKG can be performed at the physical layer by using the channel fading as a source of common randomness 
\cite{SKG_implementation_survey}. An important step in SKG is the reconciliation of correlated observations 
\cite{coding1,coding3,coding4,coding7}. To this end, the codes employed should be able to reconcile any mismatched bits with high probability (reliability) while on the other hand they should not reveal any information about the generated key (security). Although the design of reconciliation codes for SKG has been studied for long blocklengths, respective results in the short blocklength have not been reported.

Furthermore, high precision indoor localization is at present considered a standard capability in 5G networks. Widely employed localization methods, including time-of-flight, multilateration, multiangulation, etc., are capable of high precision localization but typically require high complexity operations and measurements from multiple reference points. On the other hand, proximity detection has the lowest computational complexity, and could be easily implemented with the equipment already present in constrained IoT devices (e.g., using bluetooth low energy)~\cite{location_survey2} to estimate the distance from a single reference point to a transmitting beacon~\cite{proximity_example}.

Following from the above, in this paper we introduce a fast, multi-factor authentication protocol that uniquely combines the above PLS techniques, namely PUF authentication, SKG at the short blocklength, and, proximity estimation. The proposed solution provides mutual authentication between a mobile IoT node (Alice) and a static edge server (Bob). \textcolor{black}{Furthermore, as IoT devices store sensitive information, we incorporate a one-time alias scheme that preserves the privacy of the IoT node during the execution of the authentication procedure}. The contributions of this work are summarized in Table \ref{table:contributions};  we propose a two-phase authentication protocol between Alice and Bob, comprising an initial enrollment phase, during which the IoT device is registered in the server database, and, an authentication phase, during which the authenticity of both devices is verified through the exchange of several messages. The overall authentication protocol uniquely combines the following PLS schemes: 
\begin{enumerate}
\item SKG at the short blocklength; we study half rate reconciliation code design for blocklengths $n=512, 1024$ bits, as the means to fast key agreement. In particular, we implement and compare the performance of three families of Slepian Wolf (SW) reconciliation encoders, bench-marked against a known information theoretical limit \cite{SKG_upperbound}.
\item A novel, mobility-based proximity estimation is proposed as an initial authentication factor. Instead of using high complexity localization techniques, which require measurements from multiple reference points, we propose to leverage node mobility and perform successive received signal strength indicator (RSSI) measurements from multiple locations using a single device. As a proof of concept for the proposed mobility-based proximity verification we provide experimental results for two different real-life environments.
    
\item We propose to use PUFs in a zero round trip time protocol, in which, resumption keys are generated using SKG. The combination of PUFs, SKG and mobility based-proximity detection ensures security properties such as  untraceability, anonymity, protection against impersonation attacks and many more.
\item The security properties of the proposed protocol are verified through formal methods:  Burrows, Abadi, Needham (BAN)~\cite{Ban_logic} as well as  Mao  and  Boyd  (MB) logic and the  Tamarin-prover~\cite{Tamarin-prover}. To the best of our knowledge, it is the first time formal methods are used for PLS protocols.
\end{enumerate}

The rest of the paper is organized as follows: Section \ref{sec:related_work} discusses related work, Section   \ref{sec:SKG_short} discusses the  performance of different short blocklength SKG reconciliation encoders. Section \ref{sec:proximity} introduces our mobility-based proximity estimation mechanism and Section \ref{sec:authentication} presents the proposed authentication protocol whose security properties are verified in Section \ref{sec:sec_analysis}. Finally Section \ref{sec:conclusions} concludes this paper.

\section{Related Work and Contributions}
\label{sec:related_work}

Numerous PUF-based authentication protocols have been proposed, both for unilateral authentication and mutual authentication~\cite{PUF_auth_survey}. Some of the protocols assume the use of PUFs as the only factor of authentication ~\cite{PUF_forensics, Auth_protocol-sess_key-from_nonces1}. However, relying on PUFs as a single security factor can expose the system to a variety of threats, especially in an IoT scenario~\cite{PUFs_challenges}. Therefore, combining two or more independent credentials can be used to built a secure multi-factor authentication protocol~\cite{Auth_protocol_pre-shared-key, PUF_Location, Data_provenance_puf}. For example, \cite{Data_provenance_puf} proposes a privacy-preserving authentication protocol between IoT device and a server connected through a third party wireless gateway. The authors propose the use of PUFs for device authentication and RSSI measurements between the IoT device and the gateway to achieve data provenance. The process of gateway authentication is not clarified, making the scheme open to relay attacks (in the presence of a malicious gateway). In addition the authors propose a CRP update process which must be performed after each communication and therefore causing extra overhead. Moreover, the encryption of the CRP update process is performed using the same key used during authentication, opening the protocol to vulnerabilities related to key reuse.

Another multi-factor privacy preserving authentication protocol was proposed in~\cite{PUF_Location}. The proposed scheme achieves mutual authentication by combining PUFs with location information. The location estimation process uses raw RSSI measurements and its validity is confirmed through a comparison with a pre-stored threshold. However, as shown later in this paper (Fig. \ref{fig:RSSI_measurements}), raw RSSI measurements will typically vary with tens of dBms and could lead to incorrect estimation. 

A different PUF-based privacy preserving scheme was proposed in~\cite{Auth_protocol_pre-shared-key}. As a second factor of authentication the authors propose the usage of pre-shared secret keys. Unlike the above studies, this scheme takes into account the noise present in PUF structures and uses fuzzy extractors for reconciliation. However, as noted in \cite{PUF_Location}, this scheme is open to physical attacks.

To summarize, we list several similarities that were identified in earlier works~\cite{PUF_Location,  Auth_protocol_pre-shared-key,  Data_provenance_puf}: i) it is assumed that session keys are generated using pseudo random number generators (PRNG) modules that typically generate low-entropy keys and are vulnerable to many possible attacks~\cite{PRNG_prediction_attack}; ii) it is proposed to reuse the same authentication key in order to authenticate both parties, opening up key reuse vulnerability issues.

In this work, we propose a multi-factor and privacy-preserving authentication protocol entirely based on PLS techniques (described in full in Section \ref{sec:authentication}). With respect to privacy preservation, similarly to earlier works, we use a one-time alias ID scheme to preserve the privacy of the IoT device from malicious eavesdroppers. In such a scheme, the IoT device does not use its real ID during the authentication process, instead it uses a one-time alias ID which is updated  in every session. The key differences between the proposed protocol and previous works are summarized below:
\begin{itemize}
\item We account for the noise naturally present in PUF measurements and employ a fuzzy extractor (FE) as a reconciliation scheme to generate helper data. The protocol uses a unique key in order to authenticated each party, which eliminates the key reuse problems discussed earlier.
\item Next, to speed up the resumption of sessions we propose a novel SKG-based 0-RTT scheme, which overcomes the problems present in current solutions~\cite{0-RTT_attacks1}. To this end, we propose the use of SKG for the generation of resumption keys, which obviates the need of using PRNGs. 
Note that, combining a PUF with a SKG scheme is a novel approach, especially, in the context of short blocklength communication (Section \ref{sec:SKG_short}). The introduction of SKG allows users to send data on the first flight (early data) with both forward secrecy and protection against replay attacks, see Section~\ref{subsec:0rtt}. 
\item Finally, we propose a mobility based proximity estimation mechanism using a Kalman filter as a smoothing technique (Section \ref{sec:proximity}). The novelty of the proposed mechanism is that we leverage the mobility of IoT devices in order to prevent impersonation attacks, which has not been considered before. 
\end{itemize}

\section{SKG Slepian Wolf Short Blocklength Reconciliation} \label{sec:SKG_short}

\textcolor{black}{In the proposed authentication scheme, SKG is employed in the resumption part of the protocol. In this section we focus on the coding aspects of SKG. We consider SW reconciliation and compare the performance of different families of codes in the short blocklength $n=512, 1024$, which is pertinent for low latency communications and constrained devices.}

In our system model, we assume Alice and Bob generate binary sequences $Y_A, Y_B$ of length $n$ by quantizing their respective observations of the channel coefficient $H_0$, respectively. 
For simplicity, in this work we assume that a passive adversary, referred to as Eve, cannot obtain any information for the generated sequences\footnote{Due to large scale fading and path-loss, the channel coefficients between Alice, Bob and Eve are typically correlated. However, it is possible through pre-processing of the coefficients to remove correlations with the observation of Eve, e.g., see \cite{Prena13}, \cite{Zhu17}, \cite{LoRa19}. In the rest of this paper, for compactness of presentation we assume this assumption holds.}.

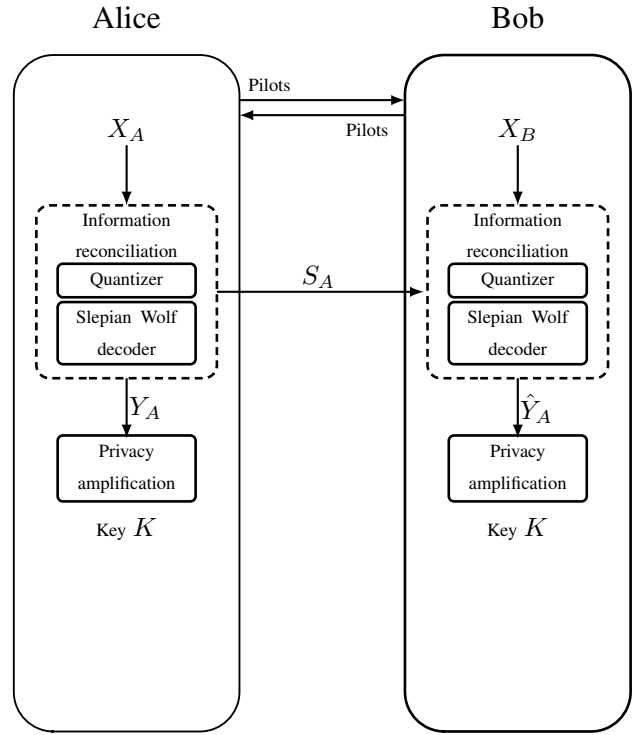
\begin{figure}
\tikzset{>=latex}
\begin{tikzpicture}
\begin{scope}[xshift=2cm]
\draw[rounded corners=15pt,line width=0.71pt] (0.5,0)--(3,0)--(3,9)--(0,9)--(0,0)--(.5,0);
\node at (1.5,9.5) {\large Alice};
\draw[thick, ->] (3,8.4) -- (5.2,8.4);
\draw[thick, ->] (5.2,8.2)--(3,8.2);
\node at (3.4,8.6) {\scriptsize Pilots};
\node at (4.7,8) {\scriptsize Pilots};
\node at (1.5,8) {${X}_A$};
\draw[thick, ->] (1.5,7.8)--(1.5,7);
\node[draw, rounded corners=2pt,line width=1pt, text width=1.6cm, align=center] at (1.5,6) {\Longstack[c]{\scriptsize Quantizer}};
\node[draw, rounded corners=2pt,line width=1pt, text width=1.6cm, align=center] at (1.5,5.3) {\Longstack[c]{\scriptsize Slepian Wolf \\ \scriptsize decoder }};
\draw[rounded corners=5pt,line width=1pt,densely dashed] (1,4.7)--(2.7,4.7)--(2.7,7)--(0.3,7)--(0.3,4.7)--(1,4.7);
\node at (1.5,6.6) {\Longstack[c]{\scriptsize Information \\ \scriptsize reconciliation}};
\draw[thick, ->] (2.7,5.85)--(5.45,5.85);
\node at (4.05,6.05) {${S}_A$};
\draw[thick, ->] (1.5,4.7)--(1.5,3.9);
\node at (1.75,4.3) {${Y}_A$};
\node[draw, rounded corners=2pt,line width=1pt, text width=1.6cm, align=center] at (1.5,3.5) {\Longstack[c]{\scriptsize Privacy \\ \scriptsize amplification }};
\node at (1.5,2.7) {\scriptsize Key \normalsize  ${K}$};
;
\end{scope}

\begin{scope}[xshift=7.2cm]
\draw[rounded corners=15pt,line width=1pt] (0.5,0)--(3,0)--(3,9)--(0,9)--(0,0)--(.5,0);
\node at (1.5,9.5) {\large Bob};
\node at (1.5,8) {${X}_B$};
\draw[thick, ->] (1.5,7.8)--(1.5,7);
\node[draw, rounded corners=2pt,line width=1pt, text width=1.6cm, align=center] at (1.5,6) {\Longstack[c]{\scriptsize Quantizer}};
\node[draw, rounded corners=2pt,line width=1pt, text width=1.6cm, align=center] at (1.5,5.3) {\Longstack[c]{\scriptsize Slepian Wolf \\ \scriptsize decoder }};
\draw[rounded corners=5pt,line width=1pt,densely dashed] (1,4.7)--(2.7,4.7)--(2.7,7)--(0.3,7)--(0.3,4.7)--(1,4.7);
\node at (1.5,6.6) {\Longstack[c]{\scriptsize Information \\ \scriptsize reconciliation}};
\draw[thick, ->] (1.5,4.7)--(1.5,3.9);
\node at (1.75,4.3) {$\hat{Y}_A$};
\node[draw, rounded corners=2pt,line width=1pt, text width=1.6cm, align=center] at (1.5,3.5) {\Longstack[c]{\scriptsize Privacy \\ \scriptsize amplification }};
\node at (1.5,2.7) {\scriptsize Key \normalsize  ${K}$};
\end{scope}
\end{tikzpicture}
\caption{System model for coding on secret key generation.}
\label{Fig:SysModel}
\end{figure}

We assume that the generated binary sequences are independent and identically distributed (i.i.d.) with equal probabilities to be $0$ or $1$, i.e., $Pr\left(Y_A[i]=0\right)=Pr\left(Y_B[i]=0\right)=0.5$, and $Pr\left(Y_A[i]\neq Y_B[i]\right) = p$ for $i=1,\ldots,n$. 
Alice and Bob aim to agree on a secret key $K$ of length $k$ that needs to be drawn uniformly from $\mathcal{K} = \{0, 1\}^k$. To this end, Alice transmits her syndrome $S_A$, of length $n-k$, through a public channel to assist Bob to obtain an estimate $\hat{Y}_A$ of her sequence $Y_A$. 
The code rate is defined as $R = \frac{k}{n}$ and the frame error rate (FER) is defined as the probability that Bob's estimation of $Y_A$ is erroneous $Pr(Y_A \neq \hat{Y}_A)$. In this set-up, Bob first estimates the sequence $\hat{Y}_A$ and then based on that, Alice and Bob independently extract the key $K$ using privacy amplification (Fig. \ref{Fig:SysModel}).

Motivated by the promising performance of low density parity check codes (LDPC) \cite{LDPCList}, polar codes \cite{PolarList} and Bose Chaudhuri Hocquenghem (BCH) codes \cite{BCHList} with list decoding in short blocklengths for standard channel coding applications, in this paper we implement and compare against the upper bound on SKG rates in \cite{SKG_upperbound}, when these three families of SW decoders are employed.

\begin{figure*} [!ht]
  \includegraphics[width=\textwidth,height=7cm]{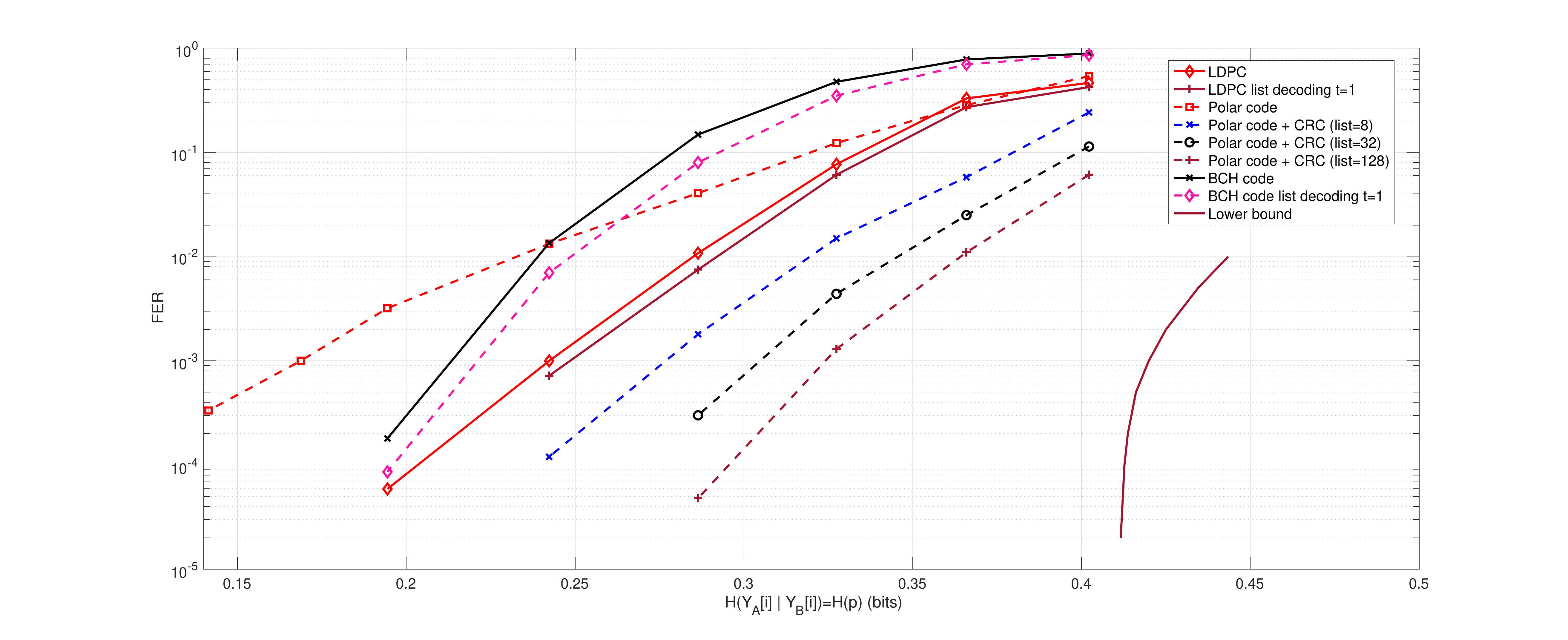}
  \caption{Comparison of FER performance of codes with the lower bound in \cite{SKG_upperbound} for $n=512$.}\label{Fig:FERN512}
\end{figure*}

\begin{figure}[!ht]
\centering
\includegraphics[ width=0.52\textwidth]{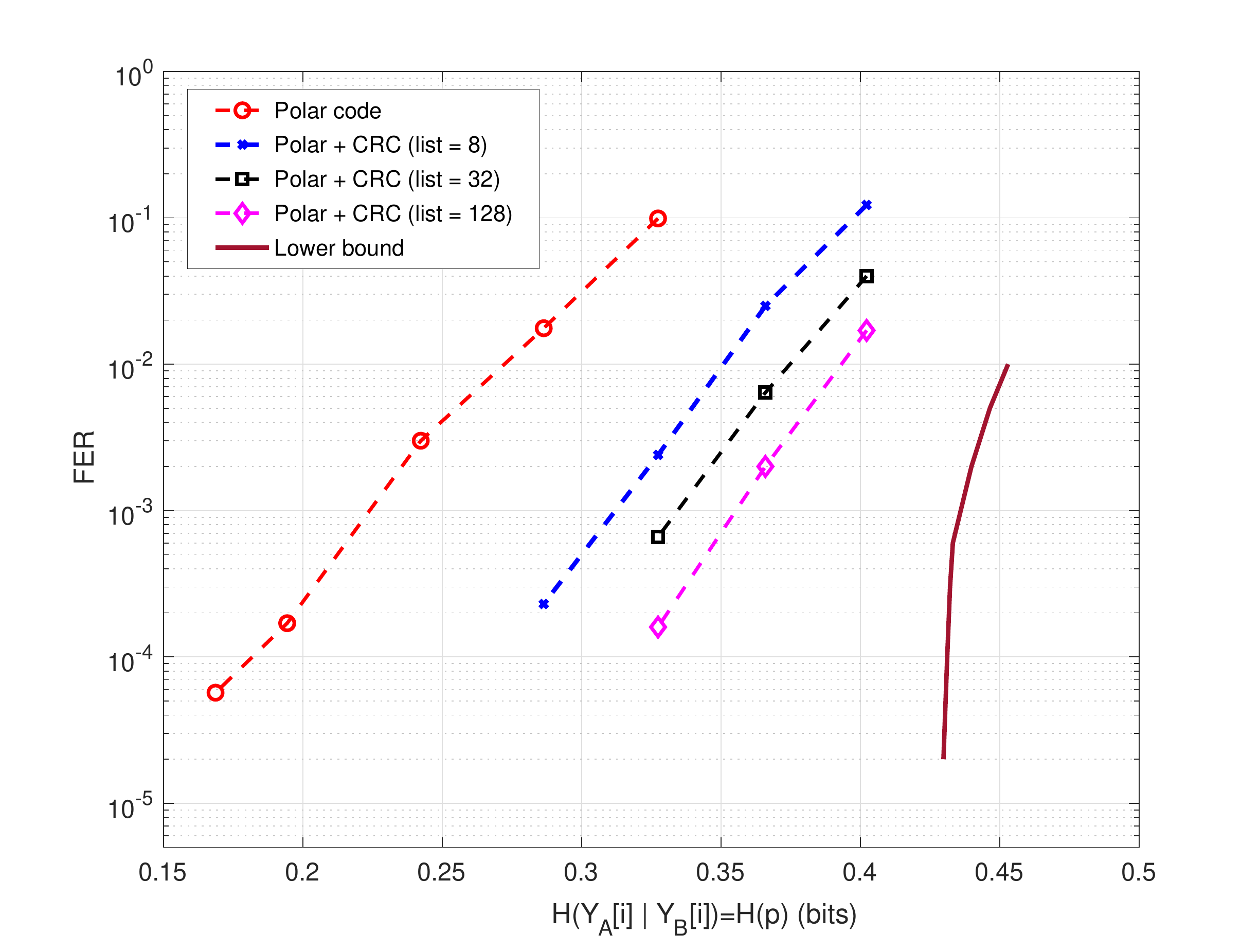}
\caption{Comparison of FER performance of codes with the lower bound in \cite{SKG_upperbound} for $n=1024$.}
\label{Fig:FERN1024}
\end{figure} 

\subsection{LDPC codes with ordered statistic decoding}
LDPC codes are powerful error correcting codes that can approach the Shannon limit at very large blocklengths. However, in general, LDPC codes do not perform well in short blocklengths.
To address such shortcomings, LDPC codes enhanced with ordered statistic decoding (OSD) is one of the techniques that has been suggested to achieve near maximum likelihood (ML) performance for short blocklengths \cite{LDPCList}.
The central idea behind  OSD is that first we pick the $\tilde k$ most reliable independent positions, where $\tilde k$ is the rank of the code. Then, based on the log-likelihood ratios (LLR) we make hard decisions on the value of the selected bits. Subsequently, we generate a candidate list of codewords by flipping the values of up to $t$ bits among them. Finally, by performing ML search in the list we choose the most likely codeword \cite{OSD2}. The size of the candidate OSD list increases with respect to $t$ as $\sum^t_{i=0} \binom ki$. In our implementation, Alice sends her syndrome $S_A = Y_A H^t$, where $H^t$ is the transpose of the parity check matrix. At the receiver side, Bob first feeds $Y_B$ and $S_A$ to the LDPC decoder to generate soft information of the bits (i.e., the LLRs). Then the LLRs are passed to the OSD block to estimate $\hat{Y}_A$.

\subsection{Polar codes with list decoding}
Polar codes are linear block error correcting codes that can provably achieve the capacity of a binary-input discrete memoryless channel as the code length tends to infinity. A significant improvement in the performance of polar codes in finite blocklengths can be achieved by utilizing successive cancellation list decoding, that keeps a list of most likely decoding paths. List decoding can be improved further by utilizing  cyclic redundancy check (CRC). The CRC assists the decoder to pick the correct decoding path in the list, even if it is not the most probable one \cite{PolarList}.

In our implementation (similar to \cite{Polar_source}), Alice encodes her sequence $Y_A$ as $U = Y_A G_n$, where $G_n = \big(\begin{smallmatrix}
  1 & 0\\
  1 & 1
\end{smallmatrix}\big)^{\otimes n}$ is the encoder matrix as defined in \cite{Polar_source}. Alice sends the syndrome $S_A$ which contains $S_1$ and CRC bits with length $l$. $S_1$ has length $n-k-l$ and contains high-entropy bits of $U$ as follows
\begin{equation}\label{eq:Polar_encoder}
H(U[i]|Y_A,U^{i-1})\geq H(U[j]|Y_A,U^{j-1}),\; 1\leq i,j\leq n,
\end{equation}
where $i$ is the position of transmitted bits and $H(\cdot)$ denotes entropy. Therefore, the actual rate of the  polar code is $R=\frac{k+l}{n}$. On the other side, Bob applies CRC-aided successive cancellation list decoding to estimate $\hat Y_A$. Note that the complexity of list decoding polar coding grows linearly with the list size. 


\subsection{BCH codes with list decoding}
BCH codes are a class of cyclic error-correcting codes constructed by polynomials over a finite field. One of the main features of BCH codes during code design is the number of guaranteed correctable error bits. A binary BCH code is defined by $(n_{BCH} ,k_{BCH} ,t_{BCH} )$, where $n_{BCH}=2^w-1$ is the blocklength, $k_{BCH}$ is the message length, and $t_{BCH}$ represents the number of guaranteed correctable error bits. To improve their error correcting capability, BCH codes can be armed with list decoding.

In our implementation of list decoding, Alice calculates the syndromes as $S_A = Y_A H^t$, where $H$ is the parity check matrix of the BCH code, and transmits it through the public channel. On the other side, Bob, first generates a candidate list by flipping up to $t$ bits of the measured sequence $Y_B$. After feeding the list to the BCH decoder, it picks the solution which is the most likely codeword with the measured sequence $Y_B$. In our implementation of list decoding, the size of the list increases with respect to $t$ as $\sum^t_{i=0} \binom ni$.

\subsection{Numerical results}
In this subsection, we analyse the FER performance of the aforementioned codes in the SKG setting and compare them with the finite length upper bound on SKG rates reported in \cite{SKG_upperbound}, which translates to a lower bound in the FER.

For instance, to generate keys that can be used with standard block ciphers, e.g., AES-128 or AES-256, the SKG process is assumed to run with blocklength 512 and rate half. We pick a regular $(3,6)$ LDPC cod with blocklength 512 bits\footnote{We note that the performance of LDPC codes can be improved by optimizing the variable and check nodes degree distributions \cite{Mahdi_LDPCdesign}, however the degree optimization of the LDPC codes is out of the scope of this work.}, a $(511,259,30)$ BCH code (the rate is slightly higher than half), and a half-rate polar code with $(n,k)=(512,267)$ and $11$ bits CRC. In Fig. \ref{Fig:FERN512}, the FER performances of half-rate codes with $512$ bits blocklength are depicted (i.e., the key length after privacy amplification is $256$) and compared to the lower bound reported in \cite{SKG_upperbound}. 

As it is demonstrated in Fig. \ref{Fig:FERN512}, although classical polar codes do not perform well in the short blocklength for Slepian Wolf coding, their performance can be significantly improved by arming them with list decoding. For example, the polar code with list size $128$, provides near three orders lower FER compared to the classical polar code at $H(Y_A[i]|Y_B[i])=0.2864$. However, this improvement comes at the cost of $128$ times more decoding complexity. On the other hand, we do not observe a significant improvement by using list decoding size $t = 1$ for LDPC and BCH code for this blocklength. We do not generate higher list decoding order for LDPC and BCH codes due to their high decoding complexity. 

Moreover, we consider an instance of $n=1024$. In Fig. \ref{Fig:FERN1024}, the FER performance of a half-rate polar code with $(n,k)=(1024,523)$ and $11$ bits CRC for different list sizes are shown. Fig. \ref{Fig:FERN1024} shows list decoding remarkably improves the code performance. Also, at blocklength $n=1024$, the gap between the FER of the polar code with list size 128 and the lower bound is less than the case $n=512$. We posit that one of the reasons is that the lower bound at this length is tighter than the first instance.

\section{Mobility-based Proximity Estimation}
\label{sec:proximity}
\begin{figure}[t]
\centering
\includegraphics[ width=0.3\textwidth]{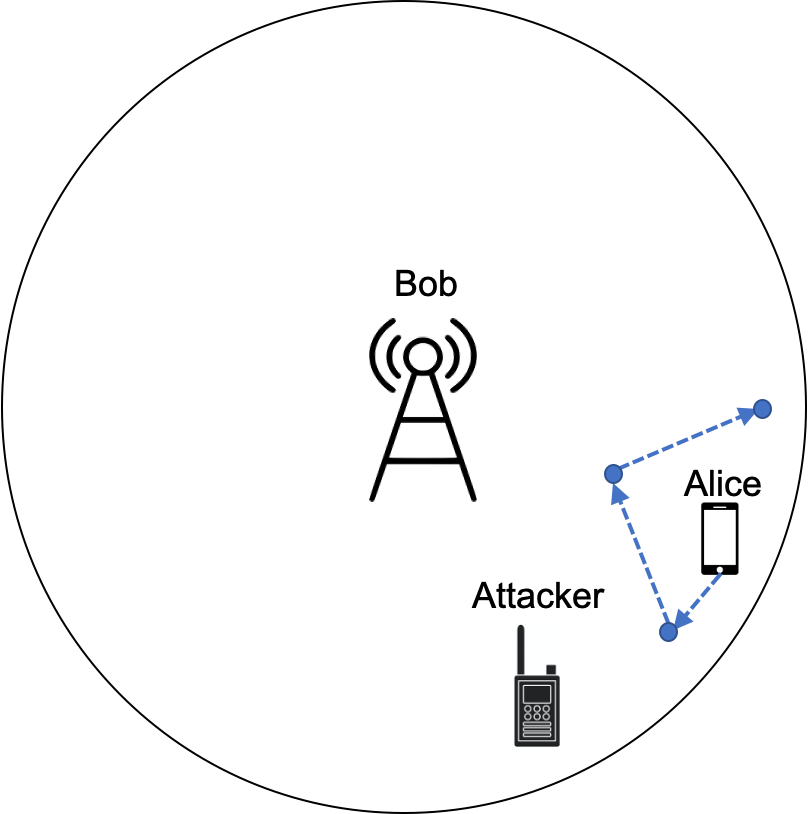}
\caption{Proposed proximity estimation.}
\label{fig: proximity}
\end{figure}

Introducing a  ``smart movement" environment brings a number of advantages to IoT systems, including energy savings, control over the node mobility and increased overall quality-of-experience (QoE)~\cite{smart_movement}. In this direction, we propose in this section a proximity estimation approach, leveraging  mobility. The novelty in our strategy relies upon the fact that if Alice (e.g., a mobile IoT node) moves in a manner unpredictable for adversaries, she can take successive measurements of the RSSI transmitted by static Bob (e.g., a static edge server) and use them for proximity estimation, as shown in Fig. \ref{fig: proximity}. In fact, this lightweight proximity estimation approach allows Alice to detect impersonation attacks\footnote{An impersonation attack when proximity estimation is used as an authentication factor can be mounted by altering the transmission power level, e.g., as in some false base station types of attack. By taking successive measurements of the RSSI in unpredictable locations, this attack can be overcome. } when used in combination with the authentication protocol presented in the next section. As a proof of concept we will present a simple implementation using Kalman filters.

Due to the ease of implementation and signal availability, RSSI-based localization is usually a favoured technique. According to the inverse-square law, the RSSI at Alice can be used to estimate the distance between her and Bob. Based on the fact that the channel coefficients follow a log-normal power distribution we assume a traditional path loss model to translate RSSI values to a   distance estimation between two nodes~\cite{rssi-to-distance2}:
\begin{equation}
    \hat{d}=d_{0}10^{\frac{P_{0}-P}{10n}}e^{-\frac{1}{2}\left(\frac{\sigma_{X_{\sigma}\ln(10)}}{10n} \right)^2}, \label{eq:distance-estimation}
\end{equation}
where $P$ is the strength of the received signal in dB, $P_{0}$ represents the average received signal strength at some reference distance $d_{0}$ in dB, $\hat{d}$ is the estimated distance to the transmitter, $n$ is an attenuation factor that gives the relation between distance and received power, and $X_{\sigma} \sim \mathcal{N}(0, \sigma^2_{X_\sigma})$ is a zero mean Gaussian random variable modelling shadowing \cite{Rappaport}.

Next, to mitigate the impact of noise present in the RSSI we employ a standard Kalman filter for the proposed proximity algorithm. Kalman filters have been widely used in literature to improve the reliability of RSSI-based localization~\cite{Domuta20}. The filter's parameters are usually in the form of matrices, however, the target in the scenario assumed here is static, and as a result all parameters reduce to scalar values. This greatly reduces the complexity of the filter and makes the algorithm suitable for a resource constrained device. The filter works by the assumption that the current state $x_i$ has a relation to the previous state $x_{i-1}$, and this relation is expressed as follows:
\begin{equation}
x_i=Ax_{i-1}+Bu_{i-1}+w_{i-1},
\end{equation}
where the transition matrix $A$ links the current state $x_i$ with the previous one $x_{i-1}$, $B$ is a control matrix which relates the control vector $u$ to the state and $w$ is i.i.d. normally distributed process noise such that $w\sim N(0,Q)$. Following that a measurement is given by 
 \begin{math}
z_i=Hx_i+R_i,
\end{math}
where $H$ is an observation matrix used to translate each state to a measurement and $R$ is i.i.d. normally distributed measurement noise such that $R\sim \mathcal{N}(0,\sigma_{R}^2)$.

In the following subsection, we provide the details of our implementation of the proposed proximity estimation using a simple Kalman filter.

\subsection{Proof of concept for mobility-based proximity estimation}

\begin{figure}[!t]
    \centering
      \hspace{-6cm} \includegraphics[clip, trim=3.5cm 12.4cm 4cm 12.5cm, width=0.48\textwidth]{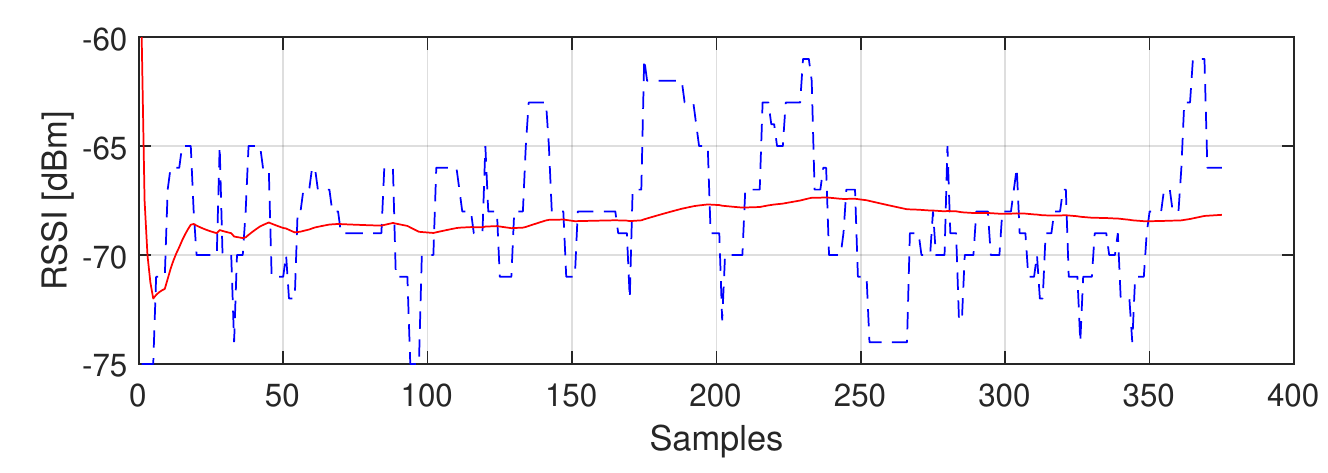}
    \hspace{-3cm}
     {\hspace{-5cm}\raisebox{2.3cm}{
      \includegraphics[clip, trim=4.95cm 20.62cm 12.75cm 8.2cm, height=0.5cm]{Measurements/all.pdf}%
    }}
    \caption{Measured RSSI data (dashed) and filtered data using Kalman filter (solid) at distance of $3$ meters. The measurement noise variance is set to $\sigma_R=0.1$.}
    \label{fig:RSSI_measurements}
    \end{figure}
We have performed a set of experiments at two different environments: i) in a small auditorium; and, ii) in a library. In both scenarios we had a static Bluetooth low energy (BLE) beacon, transmitting at $1$ dBm and a smartphone (mobile node) measuring the RSSI at different locations. This decision is motivated by the scenario assumed for this study, i.e., the access point (Bob) is fixed while the mobile IoT device (Alice) takes consecutive proximity measurements. The line of sight between the two devices was not always present due to moving people in the area. Moreover, there were other BLE and WiFi devices in the vicinity, causing further interference. 

First, in Fig. \ref{fig:RSSI_measurements}, we demonstrate the performance of the Kalman filter. The chosen  parameters are as follows: the process variance to $Q=10^{-6}$; the measurement noise variance for the specific environment was chosen as $\sigma_{R}=0.1$. and $u=0$, as no control signals are used. The results show that the filtered data quickly stabilize eliminating the noise in the measurements, while the raw RSSI data wildly fluctuate by tens of dBms.

Next, the path loss model for each environment was determined. In both scenarios the smartphone was used to measure the RSSI at different distances from the BLE beacon (1, 3, 6 meters). For each distance, we performed 50 measurements while during each measurement the mobile device collected 20 samples of the RSSI. The motivation behind this value is that in a realistic online phase, Alice could quickly (in a matter of milliseconds) collect 20 samples. Furthermore, as it can be seen in Fig. \ref{fig:RSSI_measurements} even before the 20th sample the Kalman filter has already converged and its output varies only by a few dBms. Therefore, for our proximity estimation we assume that the 20th output of the Kalman filter is the ``decision" output which Alice uses to determine her distance to Bob.

\begin{figure}[!t]
    \centering
    \includegraphics[clip, trim=3.5cm 9.4cm 4cm 10cm, width=0.48\textwidth]{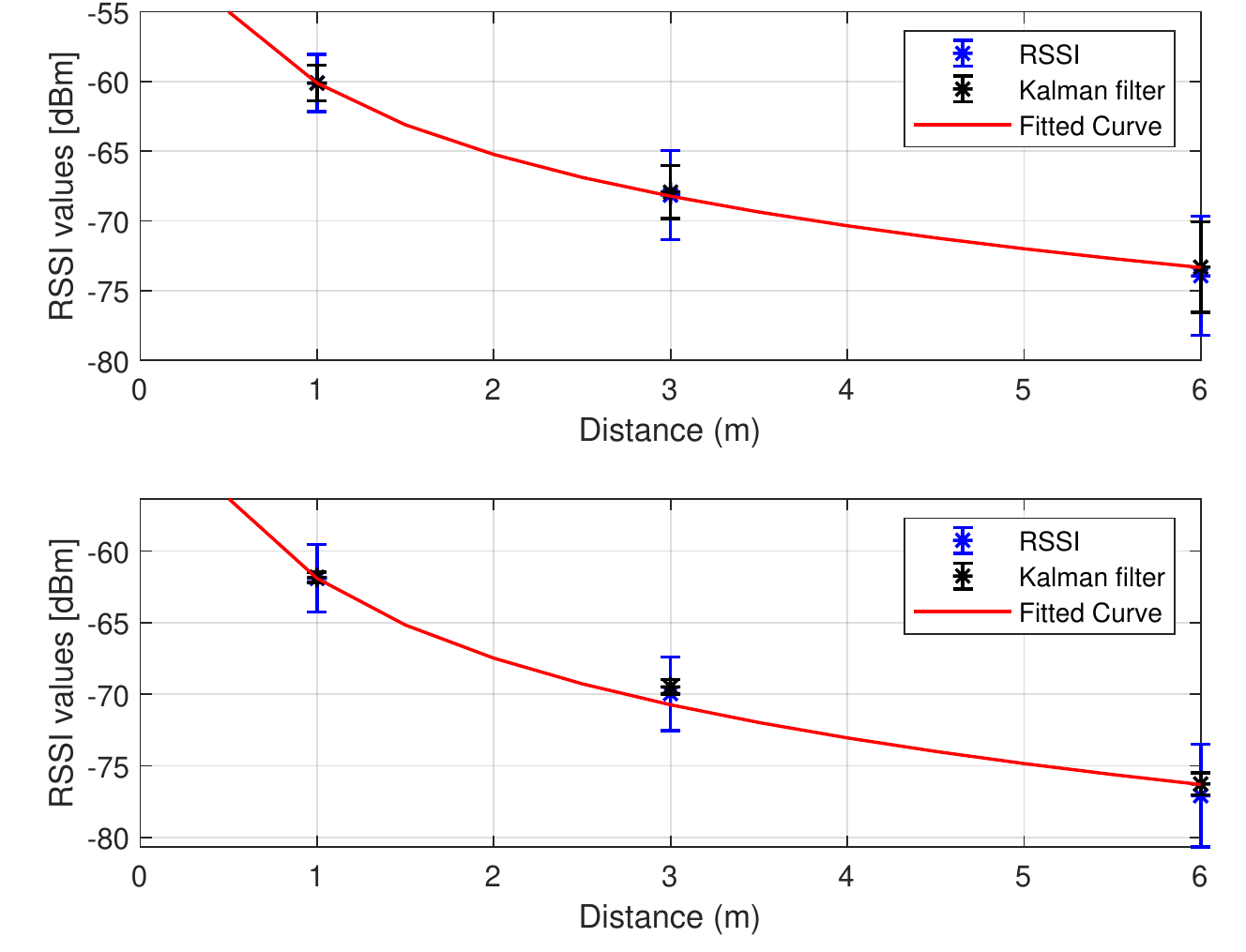}
    \caption{Curve fitting of the path loss model for a small auditorium (TOP) and a library (BOTTOM). }
    \label{fig:rssi-fit1}
\end{figure}

\begin{figure}[!t]
    \centering
    \includegraphics[clip, trim=3.5cm 9.4cm 4cm 10cm, width=0.48\textwidth]{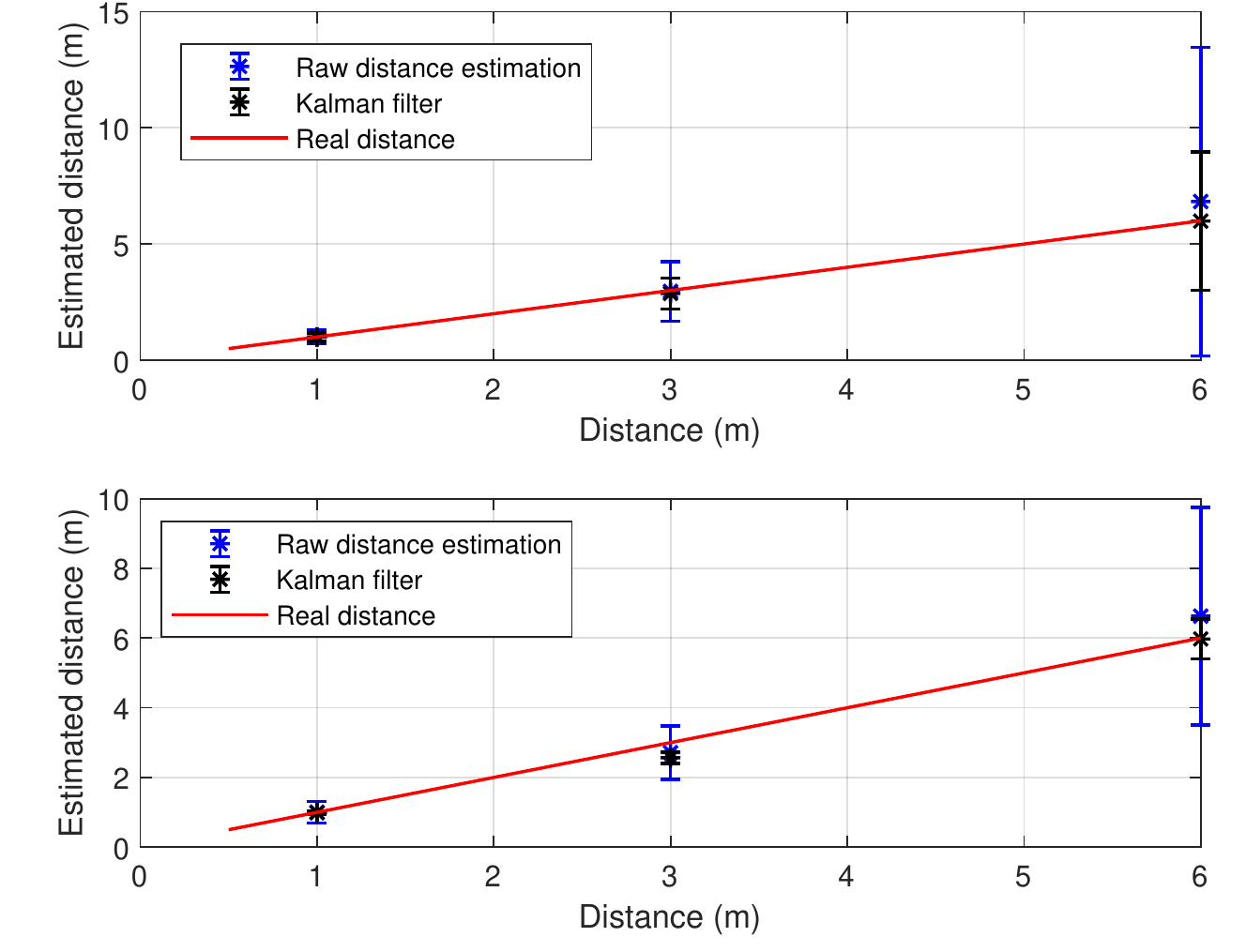}
    \caption{Distance estimation for a small auditorium (TOP) and a library (BOTTOM). }
    \label{fig:rssi-fit2}
\end{figure}

The curve fitting of the path loss model in both scenarios is given in Fig. \ref{fig:rssi-fit1}. The curves show the standard deviation of the collected RSSI data and the standard deviation of the ``decision" outputs of the Kalman filter. The estimated channel parameters for both scenarios are given in Table \ref{tab:table_parameters}. 
\begin{table}[!t]
    \centering
        \caption{Estimation of channel parameters for both scenarios }
    \label{tab:table_parameters}
    \begin{tabular}{p{2cm}|p{1.78cm}|p{1.78cm}|p{1.55cm}}
Scenario & $P_0$  & $n$ & $\sigma_{X_{\sigma}}$ \\
Small auditorium & -60.12 @ 1m & 1.7 & 6.49 \\
Library & -61.91 @ 1m & 1.85 & 6.30 
\end{tabular}
\end{table}
Finally, the distance estimations based on (\ref{eq:distance-estimation}) using the collected RSSI data and the ``decision" outputs of the Kalman filter are shown in Fig. \ref{fig:rssi-fit2}. Overall, in Figs. \ref{fig:rssi-fit1} and \ref{fig:rssi-fit2} it can be seen that using this simple mechanism greatly improves the reliability of the proposed method. It can be observed that the environmental impact over the signal, such as noise and objects, increases at greater distance. This directly influences the distribution of the RSSI and increases the variation from the mean value. However, by using the ``decision" output of the Kalman filter these variations are limited and the mobile node (Alice) can successfully determine whether the static access point (Bob) is in one of the three regions: immediate (1 m), near (3 m), or far (6 m). Moreover, since Alice moves in a manner that is unpredictable  for adversaries, a malicious node cannot impersonate Bob unless they are co-located. This simple proximity estimation technique is used as an independent factor in a multi-factor authentication protocol presented in the next section.

\section{Proposed Multi-factor Authentication Protocol}\label{sec:authentication}

This section presents a lightweight multi-factor authentication scheme, leveraging PUFs, proximity estimation and SKG. It provides a mutual authentication between Alice (a mobile IoT node) and Bob (static edge server) and consists of: an enrollment phase, an authentication phase and uses SKG as a quick resumption mechanism. We note that during the channel estimation (through pilot exchange in both directions) the parties can take measurements of the RSSI and / or of the full channel state information (CSI) if needed. 
Using the RSSI measurements Alice performs the mobility-based proximity introduced in Section \ref{sec:proximity}. She positions herself in diverse (unpredictable) locations and takes multiple measurements in order to estimate Bob's location. Next, using the CSI both Alice and Bob perform the SKG approach given in Sec. \ref{sec:SKG_short}. Before providing the overall security analysis we first present all individual primitives. The notation used throughout this section is defined as follows:
\begin{itemize}
    \item A SKG scheme generating as outputs binary vectors ${K}$ and ${S}_A$ of sizes $k=|{K}|$ and $|{S}_A|$, respectively, with ${K}\in\mathcal{K} $ denoting the key obtained after privacy amplification and ${S}_A \in \mathcal{S}$ denoting Alice's syndrome.
\item Alice's PUF denoted by $P_A$  that generates a response $R \in \mathcal{R}$ to a challenge $Ch \in \mathcal{C}h$, i.e., $R=P_A(Ch)$. Also, a pair of fuzzy extractor  algorithms, denoted by $\texttt{Gen}: \mathcal{R} \rightarrow \mathcal{K}_R \times \mathcal{H}_R$, accepting as input the PUF response and generating as outputs the identification (fuzzy) key and helper data, with corresponding reproduce algorithm $\texttt{Rep}: \mathcal{R} \times \mathcal{H}_R \rightarrow \mathcal{K}_R$, such that:
\begin{align}
   &\texttt{Gen}(R)=(H_{R},K_{R}),\\
   &\texttt{Rep}(R',H_{R})=K_{R},
\end{align}
where $R,R' \in \mathcal{R}, K_{R} \in \mathcal{K_R}$ and $H_{R} \in \mathcal{H}_R$. Generally, the $\texttt{Rep}$ function has greater computational complexity than the $\texttt{Gen}$~\cite{Reverse_fuzzy_MRR}, therefore, in the proposed scheme we perform the more complex operation on the resourceful device rather than on a constrained IoT node.

\item A symmetric encryption algorithm, e.g., AES-256 in Galois field counter mode (GCM)\footnote{We note that using a block cipher such as AES-256 in a GCM operation allows to the use of the same key $K$ to encrypt typically Gigabytes of data.}, denoted by $\texttt{Es}: \mathcal{K}\times \mathcal{M} \rightarrow \mathcal{C_T} $ where $\mathcal{C_T}$ denotes the ciphertext space with corresponding decryption $\texttt{Ds}: \mathcal{K}\times \mathcal{C_T} \rightarrow \mathcal{M}$, i.e., 
\begin{eqnarray*}
    \texttt{Es}({K}, {M})={C},\\
    \texttt{Ds}({K}, {C})={M},
\end{eqnarray*}
for ${M}\in \mathcal{M}$, ${C}\in\mathcal{C_T}$.

\item A pair of message authentication code (MAC) algorithms, denoted by $\texttt{Sign}: \mathcal{K}\times \mathcal{M}\rightarrow \mathcal{T}$, with a corresponding verification algorithm $\texttt{Ver}: \mathcal{K}\times \mathcal{M} \times \mathcal{T} \rightarrow \{yes, no\}$:
\begin{eqnarray*}
&&\texttt{Sign} ({K}, {M})={T},\\
&&\texttt{Ver} ({K}, {M}, {T})=\left\{\begin{array}{ll}
\it{yes}, & \text{if integrity verified}\\
\it{no}, & \text{if integrity not verified}
\end{array}\right. 
\end{eqnarray*}

\item A cryptographic (irreversible) one-way hash function
$$\texttt{Hash}: \{0,1\}^{q}\rightarrow \{0,1\}^k,$$
that is used to compress the size of an input binary vector of length $q $ to a binary vector of length $k=|K|$.
\end{itemize}

In all of the previously defined functions, the insertion of an index $i-1$ denotes the value of a variable or quantity one instance earlier than its corresponding value at instance $i$, e.g., $Ch_1$ denotes the PUF challenge at instance $1$ while $Ch_2$ denotes the PUF challenge at instance $2$. \textcolor{black}{Furthermore, following from the definition of PUFs, every challenge produces a unique response and corresponding helper data and authentication keys, i.e., $P_A(Ch_1)\neq P_A(Ch_2)$ and $\texttt{Gen}(P_A(Ch_1))\neq \texttt{Gen}(P_A(Ch_2))$.} \textcolor{black}{Finally, concatenation of two binary vectors $X$ and $Y$ is denoted by $(X||Y)$}.

\begin{figure}[t]
\centering
\begin{tikzpicture}
\tikzset{>=latex}
\begin{scope}[xshift=2cm]
\draw (0.3,12.49) node[draw,anchor=west, rounded corners=13pt,line width=1pt, text width=2.3cm, align=center] (node) {\Longstack[c]{\textbf{IoT node} \\Alice ID-$A$}};

\draw[thick,dashed, <->] (1.1,11) -- (7.7,11) node[midway,above] {Secure channel} node[midway,below] {Link established};
\draw[thick,dashed, ->] (1.1,9.2) -- (7.7,9.2)
node[midway,above] {$(A||\texttt{Request})$ };
\draw[thick,dashed, <-] (1.1,6) -- (7.7,6) node[midway,above] {\Longstack[l]{$(Ch_1|| Ch_2|| A_{\texttt{ID},1}|| \mathcal{C}_{emerg}|| \mathcal{A}_{\texttt{ID},emerg})$}};
\draw[thick,dashed, ->] (1.1,1.8) -- (7.7,1.8) node[midway,above] {\Longstack[l]{$(R_2|| \mathcal{R}_{emerg}||K_{R,1}||\mathcal{K}_{R,emerg})$}};



\draw (0.3,4) node[draw, anchor=west, rounded corners=1pt,line width=1pt, text width=6.8cm, align=left] (node2) {\Longstack[l]{
$R_1=P_A(Ch_1)$\\
$R_2=P_A(Ch_2)$\\
$\mathcal{R}_{emerg}=P_A(\mathcal{C}_{emerg})$\\
$\texttt{Gen}(R_1)=(H_{R,1},K_{R,1})$\\
$\texttt{Gen}(\mathcal{R}_{emerg})=(\mathcal{H}_{\mathcal{R},emerg},\mathcal{K}_{R,emerg})$\\
\textbf{Store}: $A_{\texttt{ID},1}, K_{R,1}, \mathcal{K}_{R,emerg}, \mathcal{A}_{\texttt{ID},emerg}$
}};

\coordinate[left=0.5cm of node.south] (node00);
 \draw[thick] (node00.south)--(node00|-node2.north);
\coordinate (d1) at (1.1,0);
 \draw[thick] (d1.north)--(d1|-node2.south);

\end{scope}

\begin{scope}[xshift=8.5cm]
\draw (2,12.49) node[draw,anchor=east, rounded corners=13pt,line width=1pt, text width=2.3cm, align=center] (server) {\Longstack[c]{  \textbf{Server} \\Bob ID-$B$}};

\draw (2,7.8) node[draw, anchor=east, rounded corners=1pt,line width=1pt, text width=5cm, align=left]  (server2) {\Longstack[l]{
\textbf{Generate}: $Ch_1, Ch_2, A_{\texttt{ID},1},$\\ 
\hspace{1.6cm}$\mathcal{C}_{emerg}, \mathcal{A}_{\texttt{ID},emerg}$}};
\draw (2,0.6) node[draw, anchor=east, rounded corners=1pt,line width=1pt, text width=5cm, align=left]  (server3) {\Longstack[l]{\textbf{Store}: $A_{\texttt{ID},1}, K_{R,1}, Ch_2, R_2$,\\
$\mathcal{C}_{emerg}, \mathcal{R}_{emerge}$,\\
$\mathcal{K}_{R,emerg},\mathcal{A}_{\texttt{ID},emerg}$}};

\coordinate[right=0.5cm of server.south] (server00);
\draw[thick] (server00.south)--(server00|-server2.north);
\coordinate[right=1.85cm of server2.south] (server22);
\draw[thick] (server22.south)--(server22|-server3.north);
\end{scope}
\end{tikzpicture}
\caption{Enrollment phase}
\label{fig:enrollment}
\end{figure}
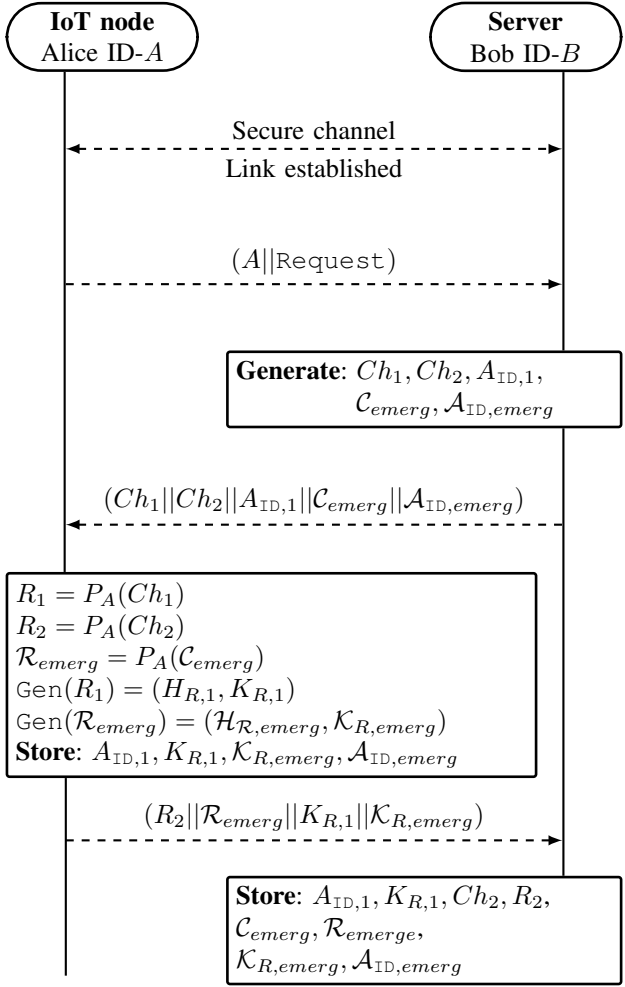

\subsection{Device enrollment}
\label{sec:enrollment}

The enrollment is a one-time operation carried out off-line over a secure channel between Alice (referred to in the following as node $A$) and Bob (referred to in the following as node $B$). The steps taken during enrollment are summarized in Fig. \ref{fig:enrollment} and are performed as follows:
\begin{enumerate}
    \item In order to establish the link between them, both devices need to exchange pilot signals. During this exchange $A$ measures the RSSI. Furthermore $A$ downloads (or creates) a map of the premises which contains the location of $B$ to enable proximity based authentication. 
    \item  After establishing the connection, Alice sends her ID $A$ with a request for registration \texttt{Request}.

    \item   Upon receiving the request, $B$ first  checks if the received ID has already been registered. If $B$ finds the ID within his database the request is rejected. If $A$ has not been registered 
    $B$ generates two initial PUF challenges $Ch_1, Ch_2$ and an initial one-time alias ID  ${A}_{\texttt{ID},1}$. These challenges will be used during subsequent authentication and will be updated with each run of the protocol. Next, $B$ generates \textit{sets} of emergency challenges and one-time alias IDs $\mathcal{C}_{emerg}$ and $\mathcal{A}_{\texttt{ID},emerg}$, respectively, such that $|\mathcal{C}_{emerg}|=|\mathcal{A}_{\texttt{ID},emerg}|$. The emergency sets are used only in a case of de-synchronisation between the devices and have multiple entries to allow for multiple recoveries. Finally, Bob sends the message $(Ch_1|| Ch_2|| A_{\texttt{ID},1}|| \mathcal{C}_{emerg}|| \mathcal{A}_{\texttt{ID},emerg})$ to Alice. Note that the two emergency sets are linked such that each element has a corresponding one in the other set.
    \item After receiving the message, Alice excites her PUF $P_A$ with $Ch_1, Ch_2$ and all challenges from the set $\mathcal{C}_{emerg}$, producing responses $R_1, R_2$ and $\mathcal{R}_{emerg}$, respectively. Next, she uses $R_1$ and $\mathcal{R}_{emerg}$ as inputs to her fuzzy extractor to generate the pair $(H_{R,1}, K_{R,1})$ and the sets of pairs $(\mathcal{H}_{R,emerg}, \mathcal{K}_{R, emerg})$. Afterwards, Alice stores ${A}_{\texttt{ID},1}, K_{R,1}, \mathcal{K}_{R, emerg}, \mathcal{A}_{\texttt{ID},emerg}$ and sends the following message to Bob $(R_2||\mathcal{R}_{emerg}|| K_{R,1}|| \mathcal{K}_{R,emerg})$.
    \item To finalise the registration process, $B$ stores the following elements that correspond to ID $A$ in his database: 
    initial authentication parameters ${A}_{\texttt{ID},1}, K_{R,1}, Ch_2, R_2$ and emergency authentication parameters in case of de-synchronisation $\mathcal{C}_{emerg}, \mathcal{R}_{emerge}, \mathcal{K}_{R,emerg}, \mathcal{A}_{\texttt{ID},emerg}$.
\end{enumerate}

\subsection{Authentication}
\label{subsec:auth_phase}
\begin{figure}[!t]
\centering
\begin{tikzpicture}
\tikzset{>=latex}
\begin{scope}[xshift=2cm]
\draw (0.3,12.49) node[draw,anchor=west, rounded corners=13pt,line width=1pt, text width=2.3cm, align=center] (node) {\Longstack[c]{\textbf{IoT node} \\Alice ID-$A$}};

\draw[thick,dashed, <->] (1.4,11) -- (7.4,11) node[midway,above] {Pilots exchange} node[midway,below] {Link established};
\draw[thick,dashed, ->] (1.4,8.2) -- (7.7,8.2)
node[midway,above] {$(A_{\texttt{ID},1}||N_1)$ };
\draw[thick,dashed, <-] (1.1,4.5) -- (7.4,4.5) node[midway,above] {\Longstack[l]{\textcolor{black}{$(C_B|| T_B)$}}};
\draw[thick,dashed, ->] (1.4,-3) -- (7.7,-3) node[midway,above] {\Longstack[l]{\textcolor{black}{$(C_A||T_A|| H_{R',2})$}}};
\draw[thick,dashed, <->] (1.4,-7.8) -- (7.4,-7.8) node[midway,above] {Secure communication} node[midway,below] {Generate $Z$};

\draw[thick,radiation,decoration={angle=45}](7.7,11) -- +(180:0.25);
\draw[thick,radiation,decoration={angle=45}](1.1,11) -- +(0:0.25);
\draw[thick,radiation,decoration={angle=45}](1.1,8.2) -- +(0:0.25);
\draw[thick,radiation,decoration={angle=45}](7.7,4.5) -- +(180:0.25);
\draw[thick,radiation,decoration={angle=45}](1.1,-3) -- +(0:0.25);
\draw[thick,radiation,decoration={angle=45}](7.7,-7.8) -- +(180:0.25);
\draw[thick,radiation,decoration={angle=45}](1.1,-7.8) -- +(0:0.25);

\draw (.3,9.7) node[draw, anchor=west, rounded corners=1pt,line width=1pt, text width=3.8cm, align=left] (node1) {\Longstack[l]{Proximity estimation\\Generate: $N_1$\\ Obtain: ${Y}_A, {S}_A, {K}$}};
\draw (.3,1) node[draw, anchor=west, rounded corners=1pt,line width=1pt, text width=6.4cm, align=left] (node2) {\Longstack[l]{
$\texttt{Ver}(K_{R1,2},C_B,T_B)$\\
$\texttt{Ds}(K_{R1,1},C_B)=(A||B||Ch_2||N_1||N_B)$\\
Generate: $N_A$\\
$R_2'=P_A(Ch_2)$\\
$\texttt{Gen}(R_2')=(H_{R',2}, K_{R',2})$\\
$Ch_{3}=\texttt{Hash}(Ch_2||N_A)$\\
$Ch_{4}=\texttt{Hash}(Ch_3||N_B)$\\
$R_3=P_A(Ch_3)$\\
$R_4=P_A(Ch_4)$\\
$\texttt{Gen}(R_3)=(H_{R,3}, K_{R,3})$\\
$A_{\texttt{ID},2}=\texttt{Hash}(A||N_B||R_{3})$\\
Break: $K_{R',2}=(K_{R'2,1},K_{R'2,2})$\\
$C_A=\texttt{Es}(K_{R'2,1},(A||B||S_A||N_A||R_{3}|| R_4))$\\
$T_A=\texttt{Sign}(K_{R'2,2}, C_A)$\\
Store: $K_{R,3}, A_{\texttt{ID},2}$
}};

\coordinate[left=0.5cm of node.south] (node00);
 \draw[thick] (node00.south)--(node00|-node1.north);
\coordinate[left=1.25cm of node1.south] (node11);
 \draw[thick] (node11.south)--(node11|-node2.north);
\coordinate (d1) at (1.1,-8.2);
 \draw[thick] (d1.north)--(d1|-node2.south);
 

\end{scope}

\begin{scope}[xshift=8.5cm]
\draw (2,12.49) node[draw,anchor=east, rounded corners=13pt,line width=1pt, text width=2.3cm, align=center] (server) {\Longstack[c]{  \textbf{Server} \\Bob ID-$B$}};

\draw (2,9.9) node[draw, anchor=east, rounded corners=1pt,line width=1pt, text width=3cm, align=left] (server1) {\Longstack[l]{
Obtain: ${Y_B}$}};
\draw (2,6.6) node[draw, anchor=east, rounded corners=1pt,line width=1pt, text width=5.8cm, align=left]  (server2) {\Longstack[l]{Verify: $A$ is at the expected distance\\
Read: CRP $(Ch_2,R_2)$ for $A_{\texttt{ID},1}$\\
Break: $K_{R,1}=(K_{R1,1}, K_{R1,2})$\\
Generate: $N_B$\\
$C_B=\texttt{Es}(K_{R1,1},(A||B||Ch_2||N_1||N_B))$\\
$T_B=\texttt{Sign}(K_{R1,2},C_B)$
}};
\draw (2,-5.2) node[draw, anchor=east, rounded corners=1pt,line width=1pt, text width=6.6cm, align=left]  (server3) {\Longstack[l]{$\texttt{Rep}(R_2, H_{R',2}) \isitequal K_{R',2}$\\
$\texttt{Ver}(K_{R'2,2}, C_A, T_A)$\\
$\texttt{Ds}(K_{R'2,1},C_A)=(A||B||S_A||N_A||R_{3}|| R_4))$\\
$Ch_3=\texttt{Hash}(Ch_2||N_A)$\\
$Ch_4=\texttt{Hash}(Ch_3||N_B)$\\
$\texttt{Gen}(R_3)=(H_{R,3},K_{R,3})$\\
$A_{\texttt{ID},2} = \texttt{Hash}(A||N_B||R_{3})$\\
\textcolor{black}{Obtain: $K$}\\
Store: $K_{R,3}, A_{\texttt{ID},2}, Ch_{4}, R_{4}$
}};

\coordinate[right=0.5cm of server.south] (server00);
\draw[thick] (server00.south)--(server00|-server1.north);
\coordinate[right=0.9cm of server1.south] (server11);
\draw[thick] (server11.south)--(server11|-server2.north);
\coordinate[right=2.25cm of server2.south] (server22);
\draw[thick] (server22.south)--(server22|-server3.north);
\coordinate (d2) at (1.22,-8.2);
 \draw[thick] (d2.north)--(d2|-server3.south);
\end{scope}
\end{tikzpicture}
\caption{Authentication protocol}
\label{fig:authentication}
\end{figure}
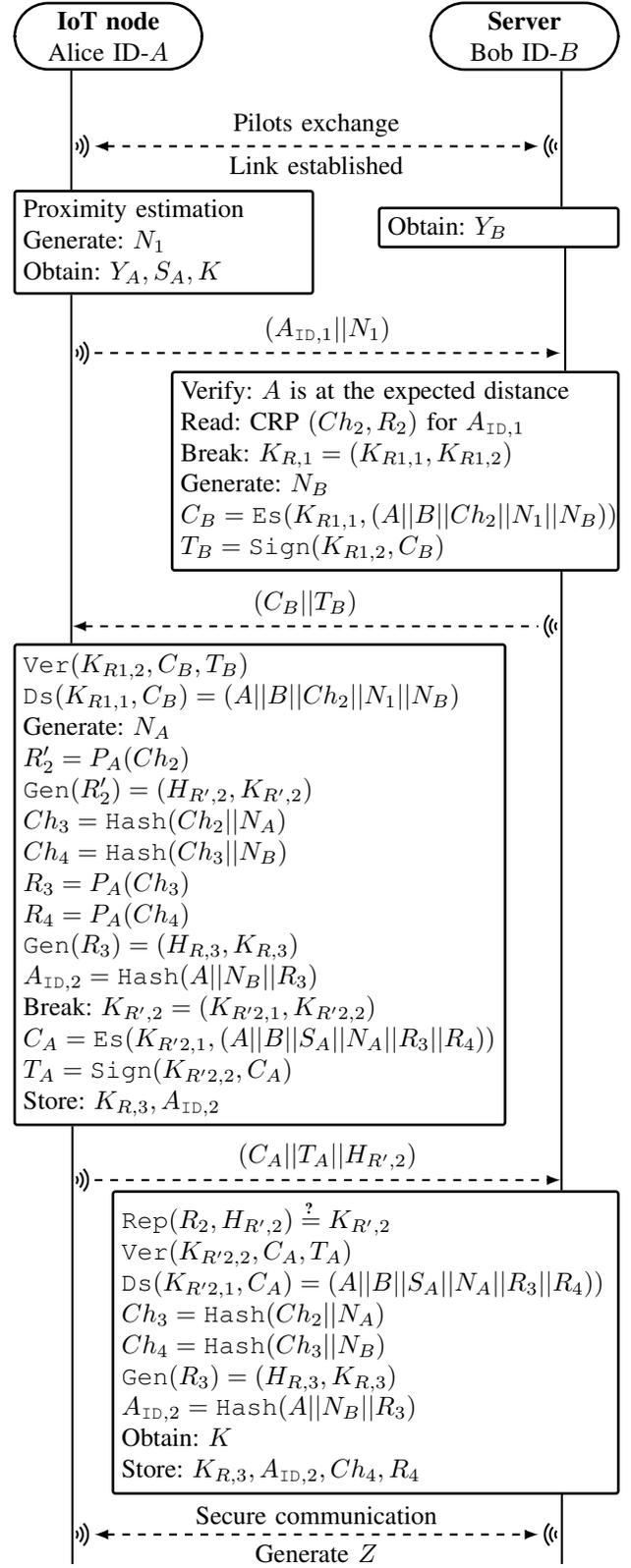

Once the enrollment is finished, both devices can use the established parameters for  future authentication over an insecure channel. The steps taken during authentication are summarized in Fig. \ref{fig:authentication} and are performed as follows:
\begin{enumerate}
    \item The devices exchange pilot signals and observe $X_A, X_B$, respectively, which they subsequently quantize to bit strings ${Y}_A$ and ${Y}_B$, correspondingly. From this step, $A$ also measures the RSSI of the received signals.
    \item Next, $A$ runs the proximity verification discussed in Section \ref{sec:proximity} to confirm the location of $B$. If the verification fails, she stops the authentication process. If it succeeds, she completes the steps of the SKG process, calculating her syndrome ${S}_A$ and key ${K}$. The key will be used later as a session key if the authentication is successful. Then, $A$ sends her request for authentication which contains a one-time alias ID $A_{\texttt{ID},i}$ and a fresh random nonce $N_1$.
    \item Upon reception, 
    $B$ accesses the database and loads the parameters that corresponds to the ID, i.e., CRP $(Ch_2, R_2)$ and key ${K}_{R,1}$. Then he generates a fresh random nonce $N_B$ and breaks $K_{R1}$ into two parts as follows: $K_{R,1}=(K_{R1,1},K_{R1,2})$. He uses the first part to encrypt $C_B=\texttt{Es}(K_{R1,1},(A||B||Ch_2||N_1||N_B))$,  and uses the second part to sign $M_B$ as: $T_B=\texttt{Sign}(K_{R1,2},C_B)$. Finally, he  sends the ciphertext $M_B$ and the signature $T_B$ to $A$.
    \item By using her stored key $K_{R,1}$, $A$ verifies the authenticity of $B$ and the integrity of the message $M_B$. If one of the verification checks fail $A$ rejects the message's claim to authenticity. If the verification succeeds she accepts and excites her PUF with the received challenge $Ch_2$. By running it on her PUF she obtains a new measurement $R'_2=P_A(Ch_2)$ and $\texttt{Gen}(R'_2)=({H}_{R',2},K_{R',2})$. Afterwards, she generates a new fresh random nonce $N_A$ and calculates the next two challenges as follows: $Ch_{3}=\texttt{Hash}(Ch_2||N_A)$ and $Ch_{4}=\texttt{Hash}(Ch_3||N_B)$. Next, she excites her PUF to produce $R_{3}$ and $R_{4}$. In order to generate the key that will be used in a future execution of the authentication protocol, $A$ executes $\texttt{Gen}(R_3)=({H}_{R,3}, K_{R,3}) $. Next, she calculates the one-time alias ID for future execution of the protocol as $A_{\texttt{ID},2}=\texttt{Hash}(A||N_B||R_{3})$ which due to the randomness of $N_B$ and $R_3$, cannot be linked to $A_{\texttt{ID},1}$. \textcolor{black}{Updating the parameter allows Alice to use a fresh ID during subsequent authentications and, therefore, preserves her privacy from eavesdroppers.} The pairs $(Ch_{4}, R_{4})$ and $(K_{R3},   A_{\texttt{ID},2})$ will be used in a subsequent connection with $B$. Next, $A$ breaks her key $K_{R',2}$ into two parts $K_{R',2}=(K_{R'2,1},K_{R'2,2})$. Similarly, to the previous step she uses half of the key to encrypt the message $C_A=\texttt{Es}(K_{R'2,1},(A||B||S_A||N_A||R_{3}|| R_4))$.
    Then, $A$ uses the second half of the key to sign the ciphertext $T_A=\texttt{Sign}(K_{R'2,2},C_A)$. Finally, $A$  sends $C_A$, $T_A$ and ${H}_{R',2}$ to $B$ and stores the pair $K_{R,3}, A_{\texttt{ID},2}$.
    \item Upon receiving the preceding message, $B$ verifies the condition $\texttt{Rep}(R_2, {H}_{R',2}) \isitequal K_{R',2}$ by using the stored $R_2$ (from the enrollment phase) and the received helper data ${H}_{R',2}$. If the verification fails, $B$ rejects the claim to authenticity. If the claim is accepted, he verifies the integrity of $C_A$ using the signed ciphertext $T_A$. Next,
    using $R_3$ and the principles of the FE, $B$ performs $\texttt{Gen}(R_3)=({H}_{R,3}, K_{R,3})$. He calculates $A_{\texttt{ID},2}=\texttt{Hash}(A||N_B||R_3)$. Following that, he stores the pairs $(K_{R,3}, A_{\texttt{ID},2}), (Ch_{4}, R_{4})$ which will be used during the next round of the protocol. Finally, using the received syndrome ${S}_A$, $B$ corrects the discrepancies in his observation ${Y}_B$ to obtain ${Y}_A$ and calculates the session key  $K=\texttt{Hash}({Y}_A)$.
    \item After the authentication process finishes, $A$ and $B$ enter the secure communication stage with session key $K$. During this stage, they generate a resumption secret ${Z}$, leveraging SKG. Instead of performing full authentication in subsequent sessions, the secret can be used as a parameter to quickly ``resume" sessions in 0-RTT. 
\end{enumerate}

\subsection{Resumption protocol}
\label{subsec:0rtt}

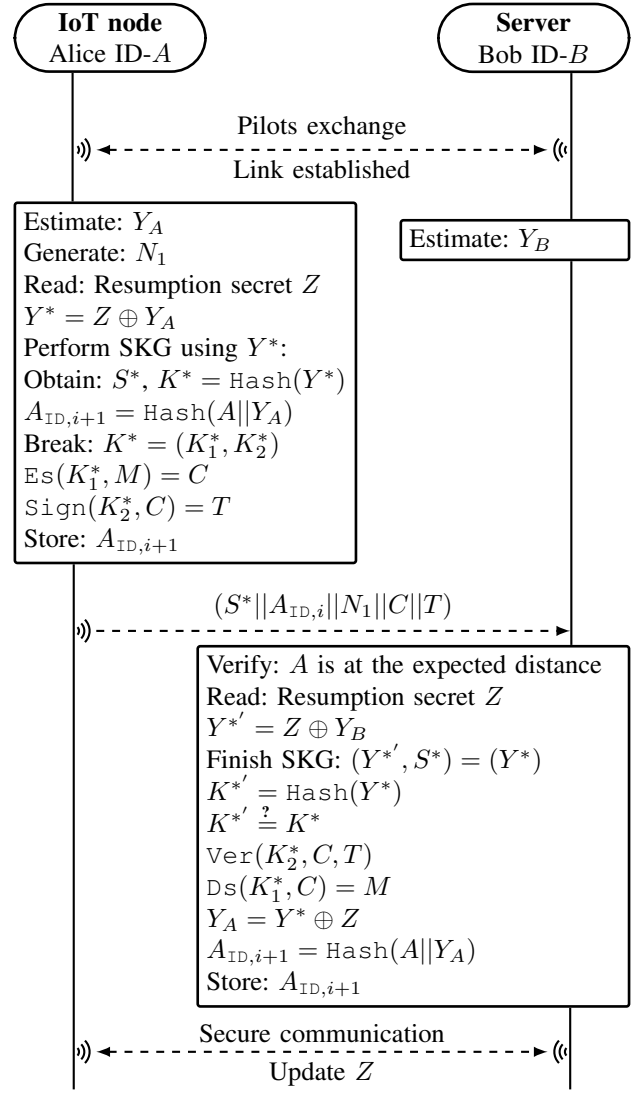
\begin{figure}[!t]
\centering
\begin{tikzpicture}
\tikzset{>=latex}
\begin{scope}[xshift=2cm]
\draw (0.3,12.49) node[draw,anchor=west, rounded corners=13pt,line width=1pt, text width=2.3cm, align=center] (node) {\Longstack[c]{\textbf{IoT node} \\Alice ID-$A$}};

\draw[thick,dashed, <->] (1.4,11) -- (7.4,11) node[midway,above] {Pilots exchange} node[midway,below] {Link established};
\draw[thick,dashed, ->] (1.4,4.6) -- (7.7,4.6)
node[midway,above] {$(S^*|| A_{\texttt{ID},i}|| N_1|| C|| T)$};
\draw[thick,dashed, <->] (1.4,-1) -- (7.4,-1) node[midway,above] {Secure communication} node[midway,below] {Update $Z$};

\draw[thick,radiation,decoration={angle=45}](7.7,11) -- +(180:0.25);
\draw[thick,radiation,decoration={angle=45}](1.1,11) -- +(0:0.25);
\draw[thick,radiation,decoration={angle=45}](1.1,4.6) -- +(0:0.25);
\draw[thick,radiation,decoration={angle=45}](7.7,-1) -- +(180:0.25);
\draw[thick,radiation,decoration={angle=45}](1.1,-1) -- +(0:0.25);

\draw (0.3,7.9) node[draw, anchor=west, rounded corners=1pt,line width=1pt, text width=4.3cm, align=left] (node1) {\Longstack[l]{ 
Estimate: $Y_A$\\
Generate: $N_1$\\ 
Read: Resumption secret $Z$\\
$Y^*=Z \oplus Y_A$\\
Perform SKG using $Y^*$:\\
Obtain: $S^*$, $K^*=\texttt{Hash}(Y^*)$\\
$A_{\texttt{ID},i+1}=\texttt{Hash}(A||Y_A)$\\
Break: $K^*=(K^*_1, K^*_2)$\\
$\texttt{Es}(K^*_1,M)=C$\\
$\texttt{Sign}(K^*_2,C)=T$\\
Store: $A_{\texttt{ID},i+1}$}};

\coordinate[left=0.5cm of node.south] (node00);
 \draw[thick] (node00.south)--(node00|-node1.north);
\coordinate (d1) at (1.1,-1.5);
 \draw[thick] (d1.north)--(d1|-node1.south);
 
\end{scope}

\begin{scope}[xshift=8.5cm]
\draw (2,12.49) node[draw,anchor=east, rounded corners=13pt,line width=1pt, text width=2.3cm, align=center] (server) {\Longstack[c]{  \textbf{Server} \\Bob ID-$B$}};

\draw (2,9.8) node[draw, anchor=east, rounded corners=1pt,line width=1pt, text width=2.8cm, align=left] (server1) {\Longstack[l]{
Estimate: $Y_B$}};
\draw (2,2) node[draw, anchor=east, rounded corners=1pt,line width=1pt, text width=5.5cm, align=left]  (server2) {\Longstack[l]{Verify: $A$ is at the expected distance\\
Read: Resumption secret $Z$\\
$Y^{*'}=Z \oplus Y_B$\\
Finish SKG: $(Y^{*'},S^*)=(Y^*)$\\
$K^{*'}=\texttt{Hash}(Y^*)$\\
$K^{*'} \isitequal K^{*}$\\
$\texttt{Ver}(K_2^*, C, T)$\\
$\texttt{Ds}(K_1^*, C)=M$\\
$Y_A=Y^* \oplus Z$\\
$A_{\texttt{ID},i+1}=\texttt{Hash}(A||Y_A)$\\
Store: $A_{\texttt{ID},i+1}$
}};

\coordinate[right=.5cm of server.south] (server00);
\draw[thick] (server00.south)--(server00|-server1.north);
\coordinate[right=0.75cm of server1.south] (server11);
\draw[thick] (server11.south)--(server11|-server2.north);
\coordinate (d2) at (1.2,-1.5);
 \draw[thick] (d2.north)--(d2|-server2.south);

\end{scope}
\end{tikzpicture}
\caption{Resumption protocol}
\label{fig:resumption}
\end{figure}

This section presents a novel physical layer resumption protocol that allows $A$ to send encrypted data in 0-RTT. During the secure communication stage of the authentication protocol in Fig. \ref{fig:authentication}, $B$ sends to $A$ a look-up identifier. Then, both derive a resumption secret $Z$ that is a function of the look-up identifier and the session parameters. 
The usage of a resumption secret for authentication helps avoid man-in-the-middle attacks in the scenario assumed here. Given the above, the resumption protocol follows the steps:

\begin{enumerate}
    \item As before, in order to establish the link both devices perform pilot exchange. $A$ and $B$ obtain  channel observations and generate sequences $Y_A$ and $Y_B$, respectively. 
    Note that, $Z$ \textcolor{black}{and  $Y_A, Y_B$ have the same length.} 
    \item 
    Next $A$, generates a fresh random nonce $N_1$ and reads the resumption secret $Z$ to generate $Y^*=Z \oplus Y_A$. Then, using her Slepian Wolf decoder she calculates the new syndrome $S^*$, that corresponds to $Y^*$, and generates the session key as $K^*=\texttt{Hash}(Y^*)$. She also calculates the one-time alias ID that will be used for subsequent session as: $A_{\texttt{ID},i+1}=\texttt{Hash}(A||Y_A)$. $A$ breaks her key into two parts $K^*=(K_1^*, K_2^*)$ and uses the first part to encrypt the early 0-RTT data $M$ as $\texttt{Es}(K^*_1,M)=C$. The second part she uses to sign the cipher text $\texttt{Sign}(K^*_2,C)=T$. Finally, she sends $(S^*|| A_{\texttt{ID},i}|| N_1|| C|| T)$. Note that the key ${K^*}$ can only be obtained if both the physical layer generated key and the resumption key are valid and this method can be shown to be forward secure~\cite{0-RTT_example}.
    \item Upon receiving the output from the last step, $B$  reads the resumption secret $Z$ and obtains $Y^{*'}=Z \oplus Y_B$. Using that and the received syndrome $S^*$, $B$ obtains $K^*=\texttt{Hash}(Y^*)$. He uses the condition $K^{*'} \isitequal K^{*}$ to verify the authenticity of $A$ and the integrity of the message. If the above succeeds he  calculates $Y_A=Y^* \oplus Z$ and stores $A_{\texttt{ID},i+1}=\texttt{Hash}(A||Y_A)$. Using the obtained key, $B$ can now decrypt the message $M$.
    \item After the resumption process finishes the two devices enter the secure communication stage using $K^*$ as a session key. During this stage, they use the channel and session properties  to  generate  new shared resumption secrets that can be used in subsequent resumptions. 
\end{enumerate}

\section{Security Analysis}\label{sec:sec_analysis}

In this section, we analyze the security of the proposed multi-factor authentication protocol illustrated in Fig.\ \ref{fig:authentication}. For the purpose of our security proofs we consider a Dolev-Yao \cite{dolev_yao} type of adversary, who has control over the wireless channel between $A$ and $B$. Furthermore: 1) the adversary can send any type of messages and queries using its knowledge gained through observation; 2) all functions and operations performed by the legitimate users  during the execution of the protocol are public except $P_A(\cdot)$ and the entire enrollment phase; and, 3) the adversary can launch denial of service (DoS) attacks and block parts of the protocol in order to de-synchronize the connection between $A$ and $B$. In terms of the SKG, for simplicity, in this work we assume a rich Rayleigh multipath environment where the adversary is more than a few wavelengths away from each of the legitimate parties and the SKG rates are given as in Section III. 

\subsection{Mutual authentication}
The proposed protocol uses a set of factors to achieve mutual authentication. It uses a mobility-based proximity estimation as a first factor of authentication. This verifies whether the server 
is at the expected distance. Next, $A$ authenticates $B$ by verifying whether the correct key is used for creating $C_B$ and $T_B$. On the other hand, $B$ authenticates $A$ by first confirming the validity of the received one-time alias ID $A_{\texttt{ID},i}$ and second by verifying whether she produced a valid response to $Ch_i$. The second condition is confirmed only if $A$ uses the correct key to generate the pair $C_A$, $T_A$.   

\subsection{Untraceability and anonymity}
During the execution of the authentication protocol, $A$ must posses a valid one-time alias ID ${A}_{\texttt{ID}}$ for each session. The one-time alias identity cannot be used twice and there is no direct relationship between subsequent IDs. Thus, no one except $B$ would know the origin of the message. Furthermore, in case of de-synchronization the device can use the set of emergency IDs $\mathcal{A}_{\texttt{ID},emerg}$. After using an emergency ID it has to be deleted from $A$'s and $B$'s memory. This approach provides privacy against eavesdroppers and ensures user's anonymity and identity untraceability properties.

\subsection{Perfect forward secrecy}
Assuming an attacker compromises $A$ and obtains all stored secrets, i.e., $(K_R, A_{\texttt{ID}})$, he cannot obtain previous keys or one-time alias IDs. First, each $K_R$ is generated using a CRP and CRPs are randomly generated and independent. Hence, by obtaining $K_{R,i}$ an adversary cannot learn $K_{R,i-1}$. Next, one-time alias IDs are generated using a one-way hash function of unique parameters for each session; if an adversary obtains $A_{\texttt{ID},i}$, he can not inverse the hash function. Furthermore, using the randomness of the wireless channel ensures that session keys are unique and independent for each session. Therefore, the proposed authentication protocol ensures the perfect forward secrecy property.

\subsection{Protection against replay attack}\label{subsec:replay_attacks}
If an adversary intercepts previous communication between $A$ and $B$, he can replay the same messages and try to pass the authentication process. In the protocol presented in Fig.\ \ref{fig:authentication} none of the parameters in the initial request are allowed to be sent twice, hence, if an attacker resends the same message to $B$ the attack will be detected and the request will be rejected. Next, if the adversary tries to re-send $C_B$ to $A$, he will be detected, since the key used to encrypt $C_B$ is changed during every session. Similarly, if the adversary tries to re-send $C_A$, he will be detected and the request will be rejected because the key used to encrypt $C_A$ is changed every session. The above shows that the proposed protocol provides resistance against replay attacks.

\subsection{Protection against impersonation attack}
A successful impersonation attack will allow the adversary to be authenticated as a legitimate user. Following from above, an adversary cannot perform a replay attack, which limits his options to perform an impersonation attack. Following from that, in order to impersonate $A$ he must generate 1) a valid one-time alias ID, and, 2) a valid ciphertext $C_A$. However, due to the unclonability properties of the PUF and the fact that the connection between a device and its PUF is secure, (i.e., system on chip) the adversary cannot generate a valid ciphertext $C_A$,  hence cannot impersonate $A$. Next, in order to impersonate $B$, the adversary must posses a valid key $K_{R,1}$ and generate a valid ciphertext $C_B$. To obtain the key an adversary must compromise $A$ (an example of such a scheme vulnerable to this attack can be found in \cite{Auth_protocol_pre-shared-key}). However, even if $A$ is compromised, the attack will be detected using the proposed proximity detection approach. This shows that our multi-factor authentication protocol provides resistance against impersonation attacks.

\subsection{Resistance to DoS attack}
To ensure security against DoS and de-syncronization attacks, the authentication protocol uses unlinkable one-time alias IDs and pairs of sets with emergency parameters $(\mathcal{C}_{emerg}, \mathcal{R}_{emerge})$ and $(\mathcal{K}_{R,emerg}, \mathcal{A}_{\texttt{ID},emerg})$. If an adversary manages to block a message from a legitimate party, such that it does not reach its intended receiver, the authentication process will stop and the used $A_{\texttt{ID},i}$ will not be updated. To overcome that $A$ can use one of her emergency IDs from the set $\mathcal{A}_{\texttt{ID},emerg}$. $B$ will then read the corresponding ${K}_{R,emerg}$ from the set $\mathcal{K}_{R,emerg}$ and use it to encrypt a message containing an emergency challenge ${C}_{emerg}$ from the set $\mathcal{C}_{emerg}$. Next, both parties can continue the authentication process as usual and setup a new one-time alias ID. In order to prevent replay attacks all used emergency parameters must be deleted from the corresponding set. This approach provides resiliency against DoS to de-synchronization attacks.

\subsection{Protection against cloning attacks}
A successful cloning attack allows the adversary to use a captured device in order to obtain secrets stored on another device. In the proposed protocol each device posses a unique pair $(K_R, A_{\texttt{ID}})$. Furthermore, all devices have unique PUFs and will produce a unique response to a challenge. Hence, the adversary cannot use secrets derived from one device in order to clone another. 
\begin{figure*}
\centering
\begin{tikzpicture}
\node at (0,3.5) { $
\begin{aligned}
\scriptstyle \cfrac{ \scriptstyle \cfrac{ \scriptstyle \cfrac{ \scriptstyle \cfrac{ \scriptstyle \cfrac{ \scriptstyle B \believes \fresh{N_B} \wedge B \sees{K_{R',2}} R_3 \mathfrak{R} N_B}{\scriptstyle B \believes \fresh{R_3}}  \wedge \frac{ \scriptstyle B \believes A \sharekey{K_{R',2}} B \wedge B \sees{K_{R',2}} R_3 }{\scriptstyle B \believes A \oncesaid{K_{R',2}} R_{3}}}{\scriptstyle B \believes A \believes A \sharekey{K_{R,2}'} B}  \wedge B \believes A \believes A^c \unavailable R_3 \wedge B \believes A \oncesaid{K_{R,2}'} R_3}{{\scriptstyle B \believes A \believes \{B \cup A \}^c \unavailable R_3} } \wedge B \believes \trusted{A}}{{\scriptstyle B \believes\{ B \cup A \}^c \unavailable R_3} } \wedge B \believes \fresh{R_3}}{ \scriptstyle B \believes A \sharekey{R_3} B}  \nonumber   
\end{aligned}$};
\draw[densely dotted] (-6.1,4.75) ellipse (0.75cm and 0.4cm);
\draw[densely dotted] (-4.2,4.75) ellipse (1cm and 0.5cm);
\draw[densely dotted] (-2.2,4.65) ellipse (0.85cm and 0.4cm);
\draw[densely dotted] (-0.55,4.65) ellipse (0.7cm and 0.4cm);
\draw[densely dotted] (1.25,3.8) ellipse (1cm and 0.4cm);
\draw[densely dotted] (4.9,3.2) ellipse (0.7cm and 0.4cm);
\node at (-7.2,4.5) { \footnotesize (R3)};
\node at (0.45,4.5) { \footnotesize (R1)};
\node at (-7.2,3.85) { \footnotesize (R5)};
\node at (-7.2,3.25) { \footnotesize (R2)};
\node at (-7.2,2.8) { \footnotesize (R6)};
\node at (-7.2,2.35) { \footnotesize (R4)};
\end{tikzpicture}
\caption{Secrecy proof in tableau format demonstrating $B$ believes $R_3$ is a good shared secret between $A$ and $B$. Initial beliefs, due to communication events are denoted with ellipses. The rules used to deduce the final goal are denoted when implied.  }
\label{fig:secrecy_proofs_ban_logic}
\end{figure*}
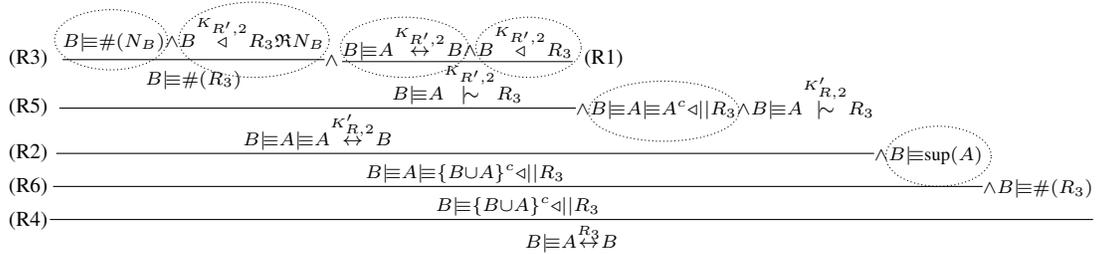

\subsection{Protection against physical attacks}
Successful physical attacks could be performed by physical tampering of the IoT device in order to change its behavior. However, by changing its behavior, the PUF will not produce the desired response and therefore $B$ will detect the attack. Therefore, the proposed protocol is resistant against physical attacks.

\subsection{Secrecy proofs using BAN and MB logic}
The secrecy evaluation of security protocols  ensures that an adversary cannot obtain or alter secret parameters. In this regards, the BAN logic~\cite{Ban_logic} is a widely used secrecy verification tool. However, some weaknesses were identified by the authors of \cite{Mao_Boyd}. They extended and improved the BAN logic to a more reliable version, namely MB logic, which is used in this paper. Formal proofs are deduced using a set of initial beliefs and rules and are based upon the message exchange within the protocol. The initial steps of MB logic are idealization of the protocol and identification of the initial beliefs. The protocol message idealization is used to interpret the implicit context-dependent information into explicit protocol specification. Based on the set of rules defined in~\cite{Mao_Boyd}, the protocol in Fig. \ref{fig:authentication} is idealised as:
\begin{enumerate}
    \item $A \rightarrow B : A_{\texttt{ID},1},N_1$
    \item $B \rightarrow A : \{N_B \mathfrak{R} N_1\}_{K_{R,1}}$
    \item  $A \rightarrow B : \{R_3 | R_4 \mathfrak{R} N_A {\mathfrak{R}} N_B \}_{K_{R,2}}$
\end{enumerate}
where $\mathfrak{R}$ gives the relation of the parameters, as defined in~\cite{Mao_Boyd}. Next, denoting  principals as $A, B$, messages and keys as $M, K$, respectively and formulas as $X$, the main properties of MB logic are: $A \believes X$ denotes $A$ believes $X$ is true; $A \sees{K} M$ denotes $A$ sees $M$ using key $K$; $A \oncesaid{K} M$ denotes $A$ encrypts $M$ using key $K$; $\fresh{M}$ denotes $M$ is of type fresh; $A \sharekey{K} B$ denotes $K$ is a good shared key between $A$ and $B$; $A \unavailable M$ denotes $M$ is not available to $A$; $\trusted{B}$ denotes $B$ is a super-principal. Following that, the inference rules defined in \cite{Mao_Boyd} and used in this paper are given in Table \ref{table_logic_rules} (Note, $\{\cdot\}^C$ denotes complement).
\renewcommand{\arraystretch}{2}
\begin{table}
\caption{Inference rules adopted from MB logic}
\label{table_logic_rules}
\resizebox{0.48\textwidth}{!}{
\begin{tabular}{ |c|l|  }
\hline
 \multicolumn{1}{|l|}{\textbf{Notation}} & \textbf{Description}\\ [0cm] 
 \hline
$\frac{A \believes A \sharekey{K} B \wedge A \sees{K} M }{A \believes B \oncesaid{K} M}$ & Authentication rule (R1)\\ [0.2cm]
 \hline
 $\frac{A  \believes \!\! B \oncesaid{K} M \wedge A \believes B \believes A \sharekey{K} B \wedge A \believes B \believes B^C \unavailable M}{A \believes B \believes \{A \cup B \}^C \unavailable M}$ & Confidentiality rule (R2)\\[0.1cm]
\hline
 $\frac{A \believes \fresh{M} \wedge A \sees{} N \mathfrak{R} M }{A \believes \fresh{N} }$ & Fresh rule (R3)\\ [0.1cm]
 \hline
 $\frac{A \believes \{A,B\}^C \unavailable K \wedge A\believes \fresh{K}}{A \believes A \sharekey{K} B}$ & Good-key rule (R4)\\ [0.2cm]
 \hline
  $\frac{A \believes \fresh{N} \wedge A \believes B \oncesaid{K} N }{A \believes B \believes A \sharekey{K} B}$ & Nonce verification rule (R5)\\[0.2cm]
  \hline
$\frac{A \believes B \believes X \wedge A \believes \trusted{B}}{ A \believes X}$ & Super-principal rule (R6)\\[0.1cm]
  \hline
\end{tabular}}
\end{table}
Given the fact that the enrollment phase is performed on a secure channel the initial beliefs can be defined  as follows:
\begin{enumerate}
    \item[A1] $A \believes A \sharekey{K_{R,1}} B$ and $B \believes A \sharekey{K_{R,1}} B$  
    \item[A2] $A \believes A \sharekey{K_{R,2}'} B$ and $B \believes A \sharekey{K_{R,2}'} B$ 
    \item[A3] $B \believes A \believes A^C \unavailable R_3 | R_4 \mathfrak{R} N_A {\mathfrak{R}} N_B$ 
    \item[A4] $B \believes \trusted{A}$ 
    \item[A5] $B \sees{K_{R,2}'} R_{3}|R_{4} \mathfrak{R} N_A \mathfrak{R} N_B$ 
    \item[A6] $A \believes B^C \unavailable R_3 | R_4 \mathfrak{R} N_A {\mathfrak{R}} N_B$ 
    \item[A7] $A \oncesaid{K_{R,2}'} R_3 | R_4 \mathfrak{R} N_A {\mathfrak{R}} N_B$
    \item[A8] $A \believes \fresh{N_1}$, $A \believes \fresh{N_A}$, $A \believes \fresh{R_3}$, $A \believes \fresh{R_4}$  
    \item[A9] $A \sees{K_{R,1}} N_B \mathfrak{R} N_1$  
    \item[A10] $A \believes B \believes B^C \unavailable N_B$ and $B \believes \fresh{N_B}$ 
    \item[A11] $A \believes \trusted{B}$ 
    \item[A12] $B \believes A^C \unavailable N_B {\mathfrak{R}} N_1$ 
    \item[A13] $B \oncesaid{K_{R,1}} N_B \mathfrak{R} N_1$ 
    \end{enumerate}

Given the initial beliefs, the authentication property of the current run of the protocol can be directly verified using the authentication rule (R1) as shown in Table \ref{table_logic_rules}. In fact, the authentication of $B$ to $A$ ($A$ to $B$) can be proven by simply using assumptions A1 and A9 (A2 and A5) in the numerator of the rule.

Next, we  prove the secrecy of parameters $R_3$ (the proofs for secrecy of $N_A$ and $R_4$ are identical) which could be used as initial belief for the next run of the protocol. The proof for $B \believes A \sharekey{R_3} B$ is given in Fig. \ref{fig:secrecy_proofs_ban_logic}. Similarly, one can prove that $A \believes A \sharekey{R_3} B$ and therefore, both parties $A$ and $B$ agree that $R_3$ is a good shared secret. However, the proof for $A$ is not presented here due to the space limitation, instead we provide a formal verification of all security properties using Tamarin-prover~\cite{Miro-Github}. Given the above and using the fuzzy extractor properties \cite{fuzzy_possible} it can be concluded that $K_{R,3}$ and $K_{R,4}$ are good shared keys between $A$ and $B$.

\subsection{Session key agreement}

It is a common practice in literature to use nonces as part of the session key generation process \cite{PUF_Location, Auth_protocol_pre-shared-key, Auth_protocol-sess_key-from_nonces1}. However, note that even if $N_A$ and $N_B$ are good shared secrets between $A$ and $B$ the low entropy of pseudo-random number generator (PRNG) modules may provoke a set of  attacks~\cite{PRNG_prediction_attack}, and lead to information leakage. Furthermore, it has been shown that true-random number generators (TRNGs) can greatly increase the time complexity in a resource limited systems making the generation time infeasible \cite{TRNG_infeasible}. Therefore, we limit the role of the nonces in the proposed scheme to only a source of freshness. On the other hand, the randomness already present in the wireless channel allows for a secure and lightweight key generation process through the SKG procedure, as illustrated in Section III. 
Finally, we note that if the session key  gets compromised, the authentication process remains secure as the adversary cannot obtain the PUF response using the session key.

\subsection{Security verification using the Tamarin-prover}
The security properties of the authentication protocol given in Section \ref{sec:authentication} were verified using the formal verification tool, Tamarin-prover~\cite{Tamarin-prover}. Tamarin was used to prove: secrecy, aliveness, weak agreement, non-injective agreement, injective agreement, untraceablity and anonymity. The model of our authentication protocol and all security proofs can be found at \cite{Miro-Github}. 

\section{Conclusions}
\label{sec:conclusions}

In this work we introduced a fast, privacy preserving, multi-factor mutual authentication protocol for IoT systems, leveraging SKG from fading coefficients, proximity estimation leveraging mobility and PUFs. 
To demonstrate the SKG performance in delay constrained applications, we provided a numerical comparison of three families of SW reconciliation codes in the short and medium blocklength regimes. 
Next, we conducted a set of experiments to demonstrate the applicability of our proposed  proximity detection in BLE networks, that leverages mobility of an IoT node. Finally, we validated the properties of the proposed authentication protocol through a detailed security analysis, using BAN and MB logic as well as the Tamarin-prover. Our analysis proves the potential of the proposed protocol as a lightweight, multi-factor alternative to the currently used computationally intensive authentication schemes, with a particular interest in IoT networks of constrained devices and wireless sensor networks.

\balance
\bibliographystyle{IEEEtran}

\bibliography{refs}

\end{document}